\documentclass{emulateapj}
\usepackage{apjfonts}
\newcommand{\hi}{H\,{\sc i}}
\newcommand{\himf}{H{\sc i}MF}
\newcommand{\himfs}{H{\sc i}MFs}
\newcommand{\msol}{\mbox{${\rm M}_\odot$}}
\newcommand{\hubble}{\mbox{$\rm km\, s^{-1}\, Mpc^{-1}$}}
\newcommand{\kms}{\mbox{$\rm km\, s^{-1}$}}
\newcommand{\mhi}{\mbox{$M_{\rm HI}$}}
\newcommand{\nhi}{\mbox{$N_{\rm HI}$}}
\newcommand{\mhis}{\mbox{$M^*_{\rm HI}$}}
\newcommand{\thetas}{\mbox{$\theta^*$}}

\newcommand{\vmax}{\mbox{$V_{\rm max}$}}

\shorttitle{The 1000 Brightest HIPASS Galaxies: H\,{\sc i} Mass Function
and $\Omega_{\rm HI}$ }
\shortauthors{Zwaan et al.}

\begin{document}

\title{The 1000 Brightest HIPASS Galaxies: \\H\,{\sc i} Mass Function
and $\Omega_{\rm HI}$ }

\author{M. A. Zwaan\altaffilmark{1},
L. Staveley-Smith\altaffilmark{2},
B. S. Koribalski\altaffilmark{2}, 
P. A. Henning\altaffilmark{3},  
V. A. Kilborn\altaffilmark{4},
S. D. Ryder\altaffilmark{5},
D. G. Barnes\altaffilmark{1},
R. Bhathal\altaffilmark{6},
P. J. Boyce\altaffilmark{7},
W. J. G. de~Blok\altaffilmark{8},
M. J. Disney\altaffilmark{8},
M. J. Drinkwater\altaffilmark{9},
R. D. Ekers\altaffilmark{2},
K. C. Freeman\altaffilmark{10},
B. K. Gibson\altaffilmark{11},
A. J. Green\altaffilmark{12},  
R. F. Haynes\altaffilmark{2},
H. Jerjen\altaffilmark{10},
S. Juraszek\altaffilmark{12},   
M. J. Kesteven\altaffilmark{2},
P. M. Knezek\altaffilmark{13},
R. C. Kraan-Korteweg\altaffilmark{14},
S. Mader\altaffilmark{2},
M. Marquarding\altaffilmark{2},  
M. Meyer\altaffilmark{1},   
R. F. Minchin\altaffilmark{8},
J. R. Mould\altaffilmark{15},
J. O'Brien\altaffilmark{10},  
T. Oosterloo\altaffilmark{16},
R. M. Price\altaffilmark{3},  
M. E. Putman\altaffilmark{17},  
E. Ryan-Weber\altaffilmark{1,2},
E. M. Sadler\altaffilmark{12}, 
A. Schr\"oder\altaffilmark{18},
I. M. Stewart\altaffilmark{18},
F. Stootman\altaffilmark{6},     
B. Warren\altaffilmark{10},   
M. Waugh\altaffilmark{1},   
R. L. Webster\altaffilmark{1}, 
and A. E. Wright\altaffilmark{2}
}

\altaffiltext{1}{School of Physics, University of Melbourne,  
   		 VIC 3010, Australia}
\email{mzwaan@ph.unimelb.edu.au}
\altaffiltext{2}{Australia Telescope National Facility, CSIRO, 
      		 P.O. Box 76, Epping, NSW 1710, Australia}
\altaffiltext{3}{Institute for Astrophysics, University of New Mexico,
                 800 Yale Blvd, NE, Albuquerque, NM~87131, USA.}
\altaffiltext{4}{Jodrell Bank Observatory, University of Manchester,
                 Macclesfield, Cheshire, SK11 9DL, U.K.}   
\altaffiltext{5}{Anglo-Australian Observatory,
                 P.O. Box 296, Epping, NSW~1710, Australia.}
\altaffiltext{6}{Department of Physics, University of Western Sydney Macarthur,
                 P.O. Box 555, Campbelltown, NSW~2560, Australia.}
\altaffiltext{7}{Department of Physics, University of Bristol, Tyndall
                 Ave, Bristol BS8 1TL, U.K.}
\altaffiltext{8}{Department of Physics \& Astronomy, University of Wales,
                 Cardiff, P.O. Box 913, Cardiff CF2 3YB, U.K.}
\altaffiltext{9}{Department of Physics, University of Queensland, QLD
   		 4072, Australia}
\altaffiltext{10}{Research School of Astronomy \& Astrophysics, Mount Stromlo
                 Observatory, Cotter Road, Weston, ACT~2611, Australia.}
\altaffiltext{11}{Centre for Astrophysics and Supercomputing, Swinburne
                 University of Technology, P.O. Box 218, Hawthorn, VIC 3122,
                 Australia.}
\altaffiltext{12}{School of Physics, University of Sydney,
                NSW~2006, Australia.}
\altaffiltext{13}{WIYN, Inc. 950 North Cherry Avenue, Tucson, AZ, USA.}
\altaffiltext{14}{Departamento de Astronom\'\i{a}, Universidad de Guanajuato,
                 Apartado Postal 144, Guanajuato, Gto 36000, Mexico.}
\altaffiltext{15}{National Optical Astronomy Observatories, P.O. Box
   		 26732, 950 North Cherry Avenue, Tucson, AZ, USA}
\altaffiltext{16}{ASTRON, P.O. Box 2, 7990 AA Dwingeloo, The Netherlands.}
\altaffiltext{17}{CASA, University of Colorado, Boulder, CO 80309-0389, USA.}
\altaffiltext{18}{Department of Physics \& Astronomy,
                 University of Leicester, Leicester LE1 7RH, U.K.}

\begin{abstract}
 We present a new accurate measurement of the \hi\ mass function of
galaxies from the HIPASS Bright Galaxy Catalog, a sample of 1000
galaxies with the highest \hi\ peak flux densities in the southern
($\delta<0^{\circ}$) hemisphere (Koribalski et al. 2003).  This sample
spans nearly four orders of magnitude in \hi\ mass (from $\log
(\mhi/\msol)+2\log h_{75}=6.8$ to $10.6$) and is the largest sample of
\hi\ selected galaxies to date.  We develop a bivariate maximum
likelihood technique to measure the space density of galaxies, and
show that this is a robust method, insensitive to the effects of large
scale structure.  The resulting \hi\ mass function can be fitted
satisfactorily with a Schechter function with faint-end slope
$\alpha=-1.30$.  This slope is found to be dependent on morphological
type, with later type galaxies giving steeper slopes.  We extensively
test various effects that potentially bias the determination of the
\hi\ mass function, including peculiar motions of galaxies, large
scale structure, selection bias, and inclination effects, and quantify
these biases.  The large sample of galaxies enables an accurate
measurement of the cosmological mass density of neutral gas:
$\Omega_{\rm HI}=(3.8 \pm 0.6) \times 10^{-4} h_{75}^{-1}$.  Low
surface brightness galaxies contribute only $\sim 15\%$ to this value,
consistent with previous findings.
 \end{abstract}
 

\section{Introduction}
 Estimates of the \hi\ mass function (\himf), the distribution
function of galaxies as a function of \hi\ mass, have been based on
very small samples of galaxies.  The most extensive blind \hi\ surveys
to date, the Arecibo \hi\ Strip Survey (AHISS, Zwaan et al.  1997)
and the Arecibo Dual-Beam Survey (ADBS, Rosenberg \& Schneider 2002)
resulted in 66 and 265 galaxy detections, respectively.  On the
contrary, state of the art measurements of the optical equivalent of
the \himf, the galaxy luminosity function (LF), are based on samples
of typically a few $\times 10^5$ galaxies (Folkes et al.  1999,
Norberg et al.  2001, 2dF; Blanton et al.  2001, SDSS).  These large
area, low redshift surveys have resulted in a very accurate census of
the local galaxy population and have shown that the power law
faint-end slope of the LF is `flat' ($\alpha=-1.2$) down to the lowest
luminosities ($M_B-5 \log h_{75} < -15$).  The shape of the LF and the
derived integral cosmological luminosity density $\rho_L$ provide
important constraints to galaxy evolution models.  However, for an
accurate assessment of the distribution and content of baryons in the
universe, it is important to also measure the \himf\ with high
accuracy.

The \hi\ Parkes All Sky Survey (HIPASS) is a blind survey of the whole
southern sky south of $\delta=25^\circ$ in the velocity range $-1200< cz
< 12700 \,\kms$.  Analysis of the extragalactic component of the survey
is ongoing and the final galaxy sample is estimated to have $\sim 7000$
entries.  This deep catalog is expected to make important
contributions to the mapping and understanding of large scale structure,
the interpretation of QSO absorption-lines, analysis of galaxy groups
and clusters, etc.  Here we use the first product of the HIPASS
database, the HIPASS Bright Galaxy Catalog (BGC, Koribalski et al. 
2003), which consists of the 1000 galaxies with the highest peak flux
densities.  This sample is four times as large as the next largest \hi\
selected galaxy sample (ADBS) and therefore enables a more careful
analysis of the shape of the \himf\, as well as its dependence on various
galaxy parameters. 

An accurate measurement of the \himf\ is highly relevant to several
modern astrophysical problems.  Firstly, the \himf\ provides a
measurement of the cosmological mass density of \hi, $\Omega_{\rm HI}$,
in the local universe.  This is an important benchmark in the mapping
out of the evolution of the neutral gas mass density from before the
epoch of reionization, when the vast majority of baryons was in the form
of \hi, to the present epoch where the mass in stars outweighs that in
neutral gas (Lanzetta, Wolfe \& Turnshek 1995, Storrie-Lombardi \& Wolfe
2000).  At intermediate redshifts, damped Ly-$\alpha$
systems are used to trace the evolution of $\Omega_{\rm HI}$, but
because of their low number density $dN/dz$ at low $z$, damped Ly-$\alpha$
can not be used to accurately measure $\Omega_{\rm HI}$ at $z=0$. 

Secondly, the shape of the \himf\ faint-end slope provides useful
input to galaxy formation models.  Both the LF and the \himf\ are
required to link the local galaxy population to the model prediction,
since both functions measure different baryonic components of
galaxies.  The \himf\ measures the distribution of {\em mass} of the
innate cool gas in galaxies, whereas the LF describes the distribution
of light emission of processed material (i.e.  stars), which is
non-trivially linked to its mass.  Many dwarf galaxies have large gas
fractions (e.g., Roberts \& Haynes 1994), which means that the \himf\
is closely related to the low mass end of the total mass function of
galaxies. Of course, this relation breaks down in regions of high
galaxy density, where gas poor dwarf elliptical galaxies dominate the
number counts of low mass galaxies.  Measuring the low mass end of the
mass function is specifically interesting with regard to Cold Dark
Matter (CDM) theory, which predicts an abundance of low mass objects,
which might be detectable in 21cm. However, the detectability could be
decreased if gas is ejected by early supernovae (Dekel \& Silk 1986;
Babul \& Rees 1992; Babul \& Ferguson 1996) or photoevaporized by the
cosmic UV background during reionization (Barkana \& Loeb 1999). 
Also at lower redshifts can ionization by the UV background of
lower column density regions of \hi\ disks have an effect on the slope
of the \himf\ (Corbelli, Salpeter, \& Bandiera 2001).

Recently, there has been considerable controversy over the faint-end
slope of the \himf.  Zwaan et al.  (1997) found a slope of $\alpha=-1.2$,
consistent with previous estimates based on optically selected galaxies
(Briggs \& Rao 1993).  Similar values for $\alpha$ were found by
Kraan-Korteweg et al.  (1999).  Schneider, Spitzak \& Rosenberg (1999)
report a flat \himf\ with an extremely steep low-mass slope below
$\mhi=10^8$.  In a more recent analysis Rosenberg \& Schneider (2002)
advocated a faint-end slope of $\alpha=-1.53$.  

The aim of this paper is to calculate the \himf\ with higher accuracy
using the HIPASS BGC, the largest sample of \hi\ selected galaxies 
available to date. 
Particular emphasis is directed toward understanding possible biases in
the calculation of the \himf.  In Section 2 we briefly summarize the
HIPASS specifics and describe the BGC.  In Section 3 various
estimators of the \himf\ are discussed and a new method, the
2-dimensional stepwise maximum likelihood method is described.  The \hi\
mass function is presented in Section 4, which includes a detailed
discussion on possible biases that may influence the \himf\ calculation. 
In Section 5 the results are compared to previous measurements of the
\himf.  The contribution of different galaxy types is discussed in
Section 6.  Section 7 presents a discussion on the \hi\ mass density
$\rho_{\rm HI}$, and in Section 8 the selection function of the survey is
discussed.  Finally, in Section 9 the conclusions are presented.  We use
$H_0= 75 \rm \,km\,s^{-1}Mpc^{-1}$ throughout this paper.

\newpage \section{The HIPASS Bright Galaxy Catalog}
 The BGC (Koribalski et al. 2003) is a catalog of the 1000 galaxies with
the largest \hi\ peak flux densities\footnote{In the remainder of this
paper we refer to `peak flux density' as `peak flux'.} $S_{\rm p}$ in the
southern sky ($\delta<0^{\circ}$).  The sample is based on HIPASS, for
which the observing strategy and reduction details are described in
Barnes et al.  (2001).  Here we briefly summarize the HIPASS survey
strategy. 

The observations were conducted in the period from 1997 to 2000 with
the Parkes\footnote{The Parkes telescope is part of the Australia
Telescope, which is funded by the Commonwealth of Australia for
operation as a National Facility managed by CSIRO.} 64-m radio
telescope, using the 21-cm multibeam receiver (Staveley-Smith et al.
1996).  The telescope scanned along strips of $8^\circ$ in declination
and data were recorded for thirteen independent beams, each with two
polarizations.  A total of 1024 channels over a total bandwidth of 64
MHz were recorded, resulting in a channel separation of $\Delta
v=13.2~\kms$ and a velocity resolution of $\delta v=18~\kms$ after
Tukey smoothing.  The total velocity coverage is $-1200$ to
$12700~\kms$.  After bandpass calibration, continuum subtraction and
gridding into $8^{\circ}\times 8^{\circ}$ cubes, the typical rms noise
is 13 mJy/beam.  This leads to a $3\sigma$ column density limit of
$\approx 6\times 10^{18} \rm cm^{-2}$ for gas filling the beam.  The
spatial resolution of the gridded data is $15\farcm5$.

An automatic galaxy finding algorithm (see Kilborn et al.  2002) was
applied to the HIPASS data set to identify all sources with $S_{\rm p}>
60$ mJy.  The list of potential detections was inspected by eye to
separate radio frequency interference and bandpass ripples from real
\hi\ sources.  Since the noise in the HIPASS data is considerably higher
at low Galactic latitude, the list of detections was complemented with
$|b|<3^{\circ}$ detections from the \hi\ Zone of Avoidance survey
(Henning et al.  2000).  Furthermore, to avoid confusion with the Milky
Way Galaxy and high velocity clouds, the range $|v|<350~\kms$ was
excluded from the list and substituted with known nearby galaxies.  
From the resulting list the 1000 galaxies with the highest
\hi\ peak fluxes were selected, resulting in a selection limit of
$S_{\rm p}>116$~mJy.  Further details of the selection, as well as
properties of the BGC galaxies, are given in Koribalski et al.  (2003). 
An important point to note here is that only four BGC sources 
outside the Zone of Avoidance ($b>5^\circ$) have no optical
counterpart on the Digitized Sky Survey. One is obscured by the LMC,
one is a tidal cloud of NGC 2442 (Ryder et al. 2001) and the other two
have heliocentric velocities $<500~\kms$ and are probably Magellanic
debris (Koribalski et al. 2003). 

HIPASS achieves 100\% coverage over the southern sky, and has a very
uniform noise level. However, there are regions of the sky near strong
sources of radio continuum, where the noise level is elevated. Near
such sources, it is possible that we are missing bright galaxies. As
above, the typical rms noise level is 13 mJy/beam. However, above an
rms noise level of 23 mJy/beam, the faintest BGC objects are detected
with a signal-to-noise ratio of no more than 5 in each velocity
channel. Within the BGC velocity range, 2.5\% of all southern HIPASS
spectra have an rms noise greater than 23 mJy/beam.  This gives a
useful upper limit to the incompleteness of the BGC.

\newpage
\section{Calculating the HI Mass Function}

\subsection{Methods}\label{methods.sec}
We define the \hi\ mass function $\theta(\mhi)$ as the space
density of objects in units of $h_{75}^3 \rm Mpc^{-3}$. The \himf\ is
normally calculated per decade of \hi\ mass, and plotted on a
logarithmic scale. For fitting purposes, we use the Schechter function
(Schechter, 1976) defined by 
\begin{equation}
 \theta(\mhi)d\mhi=\theta^* (\frac{\mhi}{\mhis})^{\alpha}
\exp(-\frac{\mhi}{\mhis})d\mhi, \label{schechter.eq} 
 \end{equation}
 characterized by the three
parameters $\alpha$, $\mhis$, and $\theta^*$, which define the slope of
the power-law, the \hi\ mass corresponding to the "knee", 
and the normalization, respectively. 

The $1/\vmax$ method originally developed by Schmidt (1968) to study the
evolution of quasars, simply consists of summing in \hi\ mass bins the
reciprocals of \vmax, the volume corresponding to the maximum distance
$D_{{\rm lim},i}$ at which an object can be detected.  This is a popular
method for determining \hi\ mass functions because the values of \vmax\
can be readily evaluated for every survey if the survey sensitivity is
well-understood (see Zwaan 1997, Rosenberg \& Schneider 2002, Kilborn
2000).  The method also works well for surveys that are not purely
flux-limited, but where the survey selection is a complicated function
of various galaxy parameters and telescope properties.  For the BGC,
$D_{{\rm lim},i}$ can simply be found by multiplying the distance $D_i$
at which the object is detected with $(S_i/S_{\rm lim})^{1/2}$, where
$S_i$ and $S_{\rm lim}$ are the peak flux of object $i$ and the limiting
peak flux of the sample, which is 116 mJy. 

The main concern about the \vmax\ method is that it is potentially
sensitive to the influence of large scale structure.  The \vmax\ method
intrinsically assumes that the galaxy population used to evaluate
luminosity functions or mass functions is homogeneously distributed in
space.  In sensitivity-limited samples (as opposed to volume-limited
samples), a correlation exists between the mass or luminosity of objects
and the distance at which they are preferentially found.  Underdense or
overdense regions may therefore yield under or over-representation of
objects of corresponding mass or luminosity.  Maximum likelihood
techniques (Sandage, Tamman \& Yahil 1979 [STY] and the stepwise maximum
likelihood method [SWML] Efstathiou, Ellis \& Peterson 1988) are designed to be
insensitive to density fluctuations and it is therefore important to
test whether these methods can be employed for our \hi\ selected galaxy
sample. 

The key of the maximum likelihood techniques is to find the parent
distribution $\theta$ which yields the maximum joint
probability of detecting all objects in the sample.  The probability
that a galaxy with \hi\ mass $M_{{\rm HI},i}$ be detected can be expressed
as
 \begin{equation}
 p(M_{{\rm HI},i}|D_i) = \frac{\theta(M_{{\rm HI},i})} 
	{\int_{M_{{\rm HI,lim}}(D_i)}^{\infty} \theta(\mhi) d \mhi},
 \end{equation}
 where $M_{{\rm HI,lim}(D_i)}$ is the minimal detectable \hi\ mass at
distance $D_i$ in Mpc. Put differently, $ p(M_{{\rm HI},i}|D_i)$ is
the fraction of galaxies
in the survey volume with \hi\ mass $M_{{\rm HI},i}$ that are near 
enough to be brighter than the survey detection limit.
Finding for which $\theta(\mhi)$ the product of 
the probabilities,
 \begin{equation}
 {\cal L}=\prod_{i=1}^{N_{\rm g}} p_i,
 \end{equation} is maximal, then gives the maximum likelihood solution
for the mass function. 
The disadvantage of this procedure is that it is parametric, that is, it
requires an analytical expression for $\theta$, which is usually taken to be
the Schechter function, as in Eq~\ref{schechter.eq}.  The stepwise maximum likelihood method (SWML)
is a modification of this procedure, 
and measures $\theta$ at fixed intervals of $\log \mhi$ by iteration.
This procedure does not depend on a functional form for $\theta$.  

As described above the maximum likelihood estimators require the
calculation of $M_{\rm HI,lim}(D_i)$, the minimal detectable \hi\ mass
at distance $D_i$.  For a hypothetical sample limited by integrated \hi\
flux, this parameter could be simply calculated by $M_{\rm
HI,lim}(D)=2.36 \times 10^5 D^2 \int SdV_{\rm lim}$, 
where $\int SdV_{\rm lim}$ is the integrated flux limit 
in Jy \kms\ and $D$ is the distance in Mpc.
For optical redshift surveys the analogous value of minimal
detectable luminosity $L_{\rm lim}(D)$
can be determined since optical redshift surveys are generally 
flux-limited, and maximum likelihood methods can be readily applied.  \hi\
selected samples are seldom integrated flux-limited, but instead the
detectability of signals depends on both the velocity width
$\Delta V$ and peak flux $S_{\rm p}$ (Zwaan et al.  1997, Rosenberg \&
Schneider 2000), or in the case of the BGC, solely on $S_{\rm p}$. 
Therefore, a unique relation between distance and the minimal
detectable \hi\ mass does not exist, and the standard maximum
likelihood techniques can not be employed. 

In principle, for every galaxy in the sample a parameter $M_{\rm
HI,lim}(D_i,P_i)$ can be calculated, which is the minimal detectable
\hi\ mass at distance $D_i$, with profile shape $P_i$ equal to that of
galaxy $i$.  Rosenberg \& Schneider (2002) apply this method to their ADBS
sample and find an \himf\ with faint-end slope $\alpha=-1.53$.  However,
this method 
can lead to incorrect measurements of the \himf.  The SWML essentially
consists of determining for each bin $k$ in $\log \mhi$ the volume
accessible to that \hi\ mass, by summing the reciprocals of space
densities of galaxies that could be detected in the volume out to the
maximum distance at which a galaxy in bin $k$ could be detected. 
A summation over all galaxies with $\mhi < M_{\rm
HI,lim}(D_i,P_i)$ is not the same. 

\label{test.sec} In the following we briefly demonstrate that this
implementation of the maximum likelihood technique for the BGC can result
in a severe overestimation of the faint-end slope $\alpha$.  We fill
volumes equal to that of the BGC with synthetic galaxy samples using
\himfs\ with slopes $\alpha=-1.0$ and $\alpha=-1.4$.  The galaxies follow
a general trend of $ \Delta V \propto \mhi^{1/3}$, similar to 
what is observed in the BGC, and we introduce
scatter on all galaxy properties equal to that seen in the BGC.  Next we
select from these volumes all galaxies with $S_{\rm p}>116~ \rm mJy$, and
calculate the \himf\ following the SWML method as described above.  From
these synthetic peak flux selected samples we also select integrated
flux-limited subsamples, for which $S_{\rm int}>25~ \rm Jy\, km\, s^{-1}$, and
also apply the SWML method to these samples.  We test by means of a
$V/V_{\rm max}$ test that the cutoff at $S_{\rm int}>25~ \rm Jy\, km\, s^{-1}$
provides a statistically complete subset.  Finally, the $1/V_{\rm max}$
method is applied to both the peak flux-limited samples and the
integrated flux-limited subsamples.  All simulations are performed 100
times. 

The results are presented in Figure~\ref{swmltest.fig}.  The top two
panels represent the results from the SWML method and the bottom two
panels those from the $1/V_{\rm max}$ method.  In each panel the 
solid lines show the input Schechter function, the circles the
measured \himfs\ for the peak flux limited samples and the open squares
the measured \himf\ for the integrated flux limited subsamples. The dashed
line is a Schechter function fit to the points.

\vbox{
\vspace{-0.3cm}
\begin{center}
\leavevmode
\hbox{%
\epsfxsize=\hsize
\epsfbox[20 145 570 700]{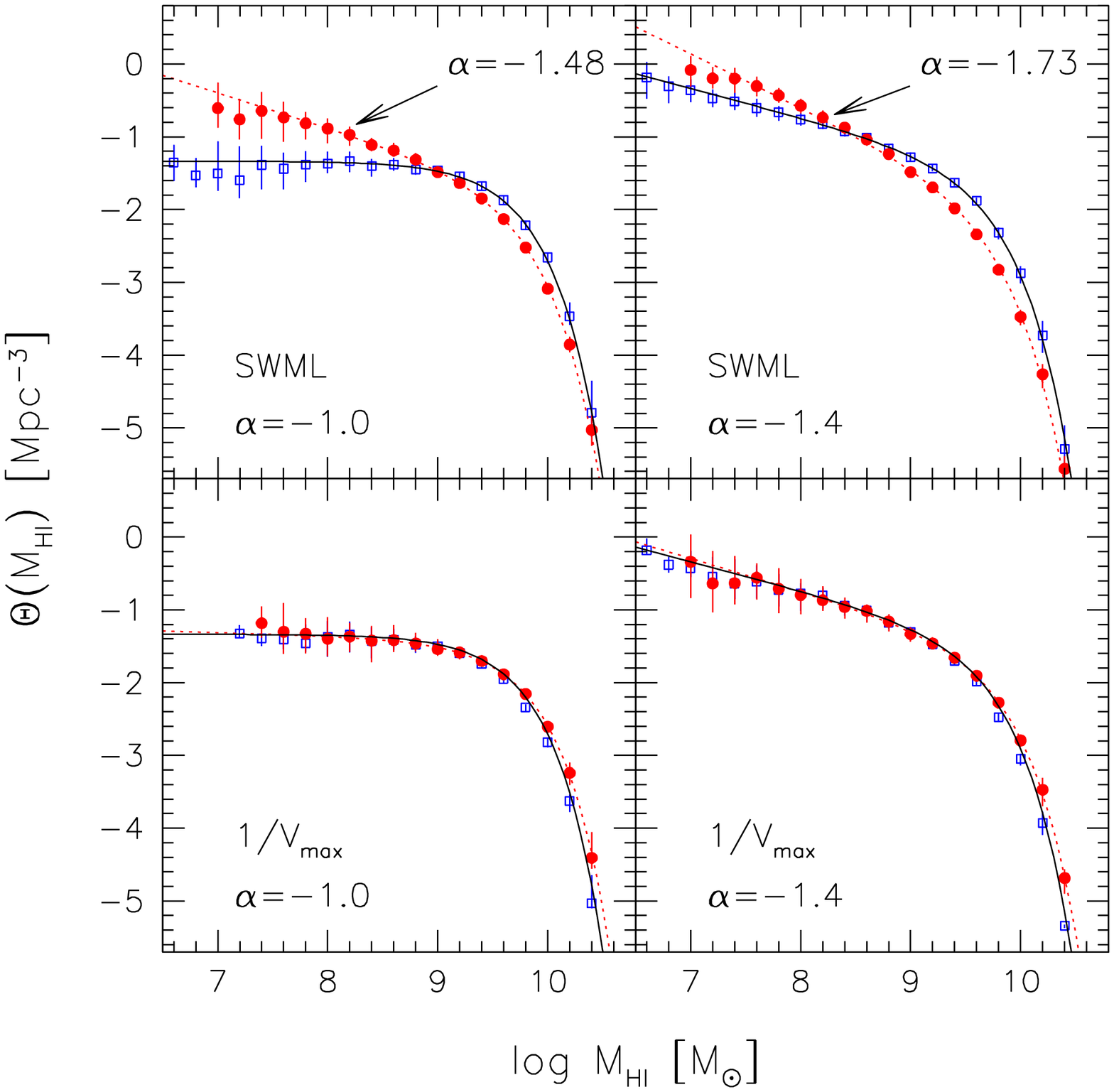}}
\figcaption{Testing the conventional SWML method on peak flux limited
samples.  In each panel the solid line shows the input Schechter
function that is used to construct synthetic galaxy samples, the filled 
circles
represent the output \hi\ mass functions of the total sample and the
squares show that of an integrated flux limited subsample.  In the top
two panels the SWML method is applied, in the bottom two the $1/V_{\rm
max}$ method is used. The SWML application on a peak flux limited sample
clearly gives incorrect results.
\label{swmltest.fig}}
\end{center}}

Both methods recover the input \himf\ very well if they are applied to the
integrated flux limited subsamples. However, this figure illustrates
clearly that the SWML method results in a severe overestimation of
the faint-end slope if it is applied to the peak flux limited samples.

We note that the examples presented here are specific to the BGC. To
which degree the output \himf\ is in error depends very strongly on
the selection function of the \hi\ survey and there are situations
possible in which the SWML coincidentally gives the correct answer.

\subsection{The 2-dimensional SWML method (2DSWML)}
 On galaxy samples that are other than flux-limited, other parameters
have to be included in the maximum likelihood calculation of the space
density of objects.  For a given \hi\ source at distance $D$, the peak
flux is directly proportional to its \hi\ mass \mhi, and inversely
proportional to the velocity width over which the flux is distributed. 
Therefore, the velocity width is the second parameter we have to include
in the maximum likelihood analysis.  A 2-dimensional maximum likelihood
algorithm can then be employed to find the true galaxy distribution. 
In the following we use $w_{20}$, the profile width measured at 20\% of
peak intensity, as a measurement of the velocity width.

The probability of detecting galaxy $i$ with \hi\ mass $M_{{\rm
HI}}^i$ and velocity width $w_{20}^i$ at distance $D_i$ is given by 
 \begin{eqnarray}
 p(M_{{\rm HI}}^i,w_{20}^i | D_i) = \hspace{4cm}\nonumber \\ \frac{\theta(M_{{\rm HI}}^i,w_{20}^i)} 
	{\int_{w_{20}=0}^{\infty}\int_{M_{\rm HI}=M_{\rm HI,lim}^i}^{\infty} 
	\theta(\mhi,w_{20}) d \mhi d w_{20}},
 \end{eqnarray}
where
 \begin{equation}
  M_{\rm HI,lim}^i(w_{20})=M_{{\rm HI}}^i\frac{S_{\rm lim}}{S_i}\frac{w_{20}}{w_{20}^i}
 \end{equation}
is the minimal \hi\ mass the galaxy could have to be detectable at
distance $D_i$ if it had a velocity width $w_{20}$,
and $\theta(\mhi,w_{20})$ is the 2-dimensional density distribution
function.
 However,  we have no
a priori knowledge of the functional form of $\theta(\mhi,w_{20})$,
which implies that a parametric maximum likelihood method 
(STY) can not be applied.

Alternatively, we can define the logarithmically binned 
2-dimensional density distribution function
 \begin{equation}
  \theta_{jk}=\theta(M,W), 
 \end{equation}
where  
\begin{equation}
 j=1,...\,,N_M\,\,\, {\rm and}\,\,\, k=1,...\,,N_W,
 \end{equation}
and $N_M$ and $N_W$ are the number of bins in $M$ and $W$, respectively,
and we define
 \begin{equation}
  M=\log (\mhi)\,\, {\rm and}\,\, W=\log(w_{20}).
 \end{equation}

The logarithm of the likelihood of detecting all galaxies in the sample 
can now be expressed as
 \begin{eqnarray}
 \ln {\cal L}=\sum_{i=1}^{N_{\rm g}}\sum_{j=1}^{N_{\rm M}}\sum_{k=1}^{N_{\rm W}}
  V(M_i-M_j,W_i-W_k) \ln \theta_{jk} \nonumber\\
  - \sum_{i=1}^{N_{\rm g}} \ln \left( 
\sum_{j=1}^{N_{\rm M}}\sum_{k=1}^{N_{\rm W}} H_{ijk} \theta_{jk}\right) +c,
 \end{eqnarray}
where $c$ is a constant and $V$ is a function defined by
 \begin{equation}
 V(x,y)=
 \left\{ \begin{array}{l}
     1 ,\, |x| \le \Delta M/2\,\,  {\rm and}\,\,  |y| \le \Delta W/2 \\
     0 ,\, {\rm otherwise} \end{array}\right. ,
 \end{equation}
which makes the first sum only go over galaxies in bin $\Delta M \Delta W$.
The function $H_{ijk}$ is defined as the fraction of the bin accessible
to source $i$:
 \begin{equation}
 H_{ijk}=\frac{1}{\Delta M \Delta W}\int_{W=W_k-\Delta W/2}^{W^+}
 \int_{M=M^{-}}^{M_j+\Delta M/2} dM dW,
 \end{equation}
where
$$
 M^{-}=\max(M_j-\Delta M/2,\min(M_j+\Delta M/2,M_{{\rm lim},i})),
$$
$$
 M_{{\rm lim},i}=M_i+\log(S_{\rm lim}/S_i)-W_i+W_k,
$$
$$
 W^{+}=\min(W_k+\Delta W/2,\max(W_k-\Delta W/2,W_{{\rm lim},i})),
$$
$$
 W_{{\rm lim},i}=W_i-\log(S_{\rm lim}/S_i)-M_i+M_j.
$$
 This implies that 
 \begin{eqnarray}
H_{ijk} = 1 \,\,\,{\rm if}\,\,\, M_i-M_j-W_i+W_k+\hspace{2cm}\nonumber\\ \log(S_{\rm
lim}/S_i)+\Delta W/2+\Delta M/2<0 \nonumber\\
H_{ijk} = 0 \,\,\,{\rm if}\,\,\, M_i-M_j-W_i+W_k+\hspace{2cm}\nonumber\\ \log(S_{\rm
lim}/S_i)-\Delta W/2-\Delta M/2>0. \nonumber
 \end{eqnarray}

Maximum likelihood solutions for $\theta_{jk}$ are now found by
differentiating ${\cal L}$ and applying the usual additional constraint
to fix the normalization (see e.g., Efstathiou et al. 1988).
We then arrive at
 \begin{eqnarray}
 \theta_{jk}=\hspace{7cm}\nonumber\\\frac{\displaystyle\sum_{i=1}^{N_{\rm g}}V(M_i-M_j,W_i-W_k)}
 {\displaystyle\sum_{i=1}^{N_{\rm g}} H_{ijk}\Delta M \Delta W 
 \left(\sum_{l=1}^{N_{\rm M}}\sum_{m=1}^{N_{\rm W}} \theta_{lm}
 H_{ilm}\Delta M \Delta W \right)^{-1} }, \label{theta.eq}
 \end{eqnarray}
and most likely values for $\theta_{jk}$ are found by iterating
Eq.~\ref{theta.eq}.  Stable solutions for $\theta_{jk}$ are usually
found after $\sim 25$ iterations. 

Finally, the \himf\ can be calculated from
 \begin{equation}
 \theta(M_j)=\sum_{k=1}^{N_W}\theta(M_j,W_k).
 \label{theta_m.eq}
 \end{equation}

In the remainder of this paper we refer to this method as the
2-dimensional stepwise maximum likelihood method or 2DSWML.  A very
similar technique was used by Loveday (2000) to calculate the $K$-band
luminosity function from a $B$-band selected galaxy sample.  The
implementation of the 2DSWML method described in this paper is
designed to work on the peak flux limited BGC, but the method could be
adjusted to work on samples other than peak flux limited.  This will
be of particular interest for the full sensitivity HIPASS galaxy
catalog (Meyer et al. 2003a) for which the selection
criteria are a combination of peak flux and velocity width.

\subsection{The selection function} \label{selec.sec}
 Whereas for the \vmax\ method the normalization of the mass function is
automatically determined, for the maximum likelihood methods the
normalization has to be determined afterwards. 
First we have to evaluate the selection function $S(D)$, which describes
the probability that an object at distance $D$ is detected by the
survey.  For a flux-limited sample the selection function would 
normally be calculated with
 \begin{equation}
 S(D)=\frac{
 \int_{{\rm max}(M_{\rm lim}(D),M_{\rm low})}^{M_{\rm high}}\theta(M) dM
 }{
 \int_{M_{\rm low}}^{M_{\rm high}}\theta(M) dM
 },
 \label{selec.eq}
 \end{equation}
 where $M_{\rm high}$ and $M_{\rm low}$ are the highest and lowest
values of $\log \mhi$ in the sample. For our 2-dimensional distribution
function this would translate to
 \begin{equation}
 S(D)=\frac{
 \int_{W_{\rm low}}^{W_{\rm high}}
 \int_{{\rm max}(M_{\rm lim}(D,W),M_{\rm low})}^{M_{\rm high}}\theta(M,W) dM dW
 }{
 \int_{W_{\rm low}}^{W_{\rm high}}
 \int_{M_{\rm low}}^{M_{\rm high}}\theta(M,W) dM dW
 }, \label{selec2d.eq}
 \end{equation} where $W_{\rm high}$ and $W_{\rm low}$ are the highest
and lowest values of $\log w_{20}$ in the sample and $M_{\rm lim}(D,W)$
is the minimal detectable \hi\ mass of a galaxy with log profile width
$W$ and at distance $D$ Mpc. The problem is that we can not derive an
analytical expression for $M_{\rm lim}(D,W)$.
It would be possible to derive an empirical relation
for $M_{\rm lim}(D,W)$, but it is far better to simply make use of the 
actual data. Therefore, we evaluate the selection function in 
Eq.~\ref{selec2d.eq} for each galaxy individually and then average the
values of $S(D_i)$ in bins of distance. 

The mean galaxy density $\bar{n}$ is then determined by correcting the
measured distance distribution of objects by the selection function.
There are various methods described in the literature of determining
$\bar{n}$ (Davis \& Huchra 1982, Willmer 1997).  Here we choose to use
the minimum-variance estimator (Davis \& Huchra 1982), but tests with
different estimators gave very similar results.  In the calculation of
$\bar{n}$, the selection function is weighted with the inverse of the
second moment of the two-point correlation function, which we set to
$J_3=8000 h_{75}^{-3} \rm Mpc^3$ (Meyer et al.  2003b).  The
value of $\bar{n}$ is found to be very insensitive to the exact value
of $J_3$.  Since at very low and high distances the number of galaxies
is small, the selection function is not accurately known in those
regions.  We therefore choose to limit the calculation of $\bar{n}$ to
$0.001<S(D)<0.1$, which roughly corresponds to the distance range
$10<D<50$ Mpc. Limiting the calculation to this distance range 
also ensures that the overdense volume within two correlation lengths
around the Milky Way Galaxy is not taken into
account in the calculation of $\bar{n}$.
Finally, we normalize the \hi\ mass function by setting
\begin{equation}
 \int_{W_{\rm low}}^{W_{\rm high}}
 \int_{M_{\rm low}}^{M_{\rm high}}\theta(M,W) dM dW
 = \bar{n}.
\end{equation}

\subsection{Testing the 2DSWML method}
 Before applying the 2-dimensional SWML method to the BGC, we run a
number of simulations to test whether the method produces reliable
results.  Similarly to the simulations in Section~\ref{test.sec} we fill
volumes with synthetic galaxy samples and select sources from the volume
in the same manner as the BGC is selected.  The galaxy samples are
constructed such that the distribution and the correlation 
statistics of peak flux, velocity
width and \hi\ mass are similar to those of the BGC and the samples
typically contain $\sim 1000$ galaxies.  Each test is based on 100
simulations.

\vbox{
\vspace{-0.1cm}
\begin{center}
\leavevmode
\hbox{%
\epsfxsize=\hsize
\epsfbox[48 160  583 700]{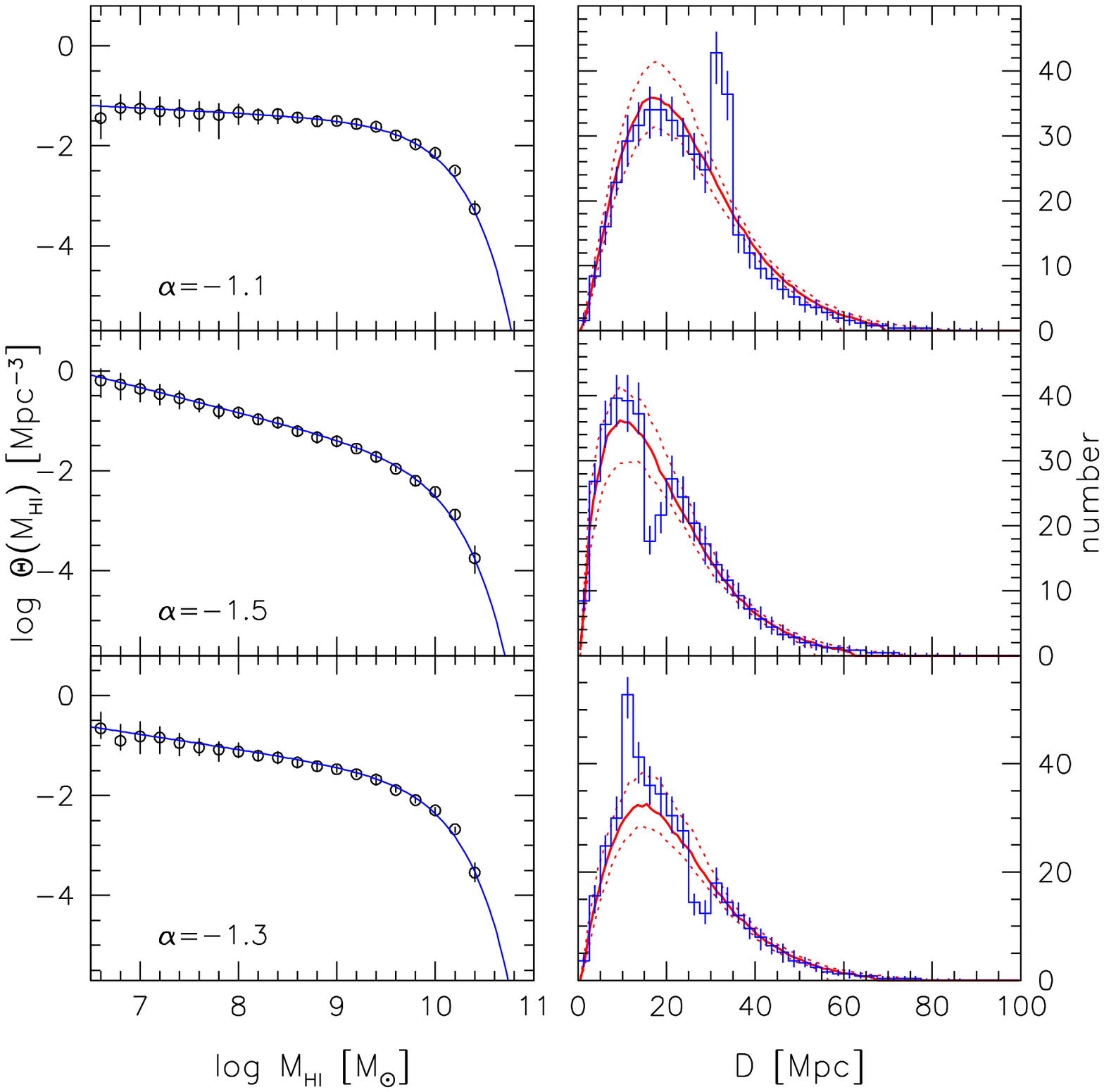}}
\figcaption{Testing the 2DSWML method.
The left panels show the input \himfs\ represented by solid lines and
recovered \himfs\ by open symbols and errorbars.  In the
right panels the thin-lined histograms show the redshift distribution of
the simulated data and the solid lines are the reconstructed selection
functions multiplied by $\Omega D^2 \Delta D \bar{n}$.  The thin dashed
lines represent $1\sigma$ uncertainties on the selection functions. 
 \label{2dtests.fig}}
 \end{center}}

Figure~\ref{2dtests.fig} shows the results of three different tests.
In the left panels the input \himfs\ are represented by solid lines,
whereas the measured \himfs\ are shown by open symbols and errorbars.
In the right panels the thin-lined histograms show the redshift
distribution of the simulated data and the solid lines are the
reconstructed selection functions multiplied by $\Omega D^2 \Delta D
\bar{n}$.  The thin dashed lines represent $1\sigma$ uncertainties on
the selection functions.  From top to bottom we show simulated data
based on 1) a faint-end slope $\alpha=-1.1$ and an overdensity at
$\sim 30$ Mpc, 2) $\alpha=-1.5$ and an underdensity at $\sim 20$ Mpc,
and 3) $\alpha=-1.3$ and an underdensity at $\sim 30$ Mpc and an
overdensity at $\sim 10$ Mpc.

It is clear that the 2DSWML method is capable of recovering the input
\himf\ with high accuracy, independent of strong over- and underdensities
in the redshift distribution of galaxies.  Also our implementation of
the calculation of the selection function gives satisfactory results. 
The selection function corresponds well with the overall redshift
distribution and is insensitive to strong density variations. 
Motivated by the success of the 2DSWML method on synthetic data, we
choose to apply it to BGC.  The results are described in the next
section. 

\section{Results}\label{results.sec}
 A greyscale representation of the density distribution function
$\theta_{jk}$ is shown in Figure~\ref{2d.fig}.  This function is
calculated with bin sizes $\Delta M=0.2$ and $\Delta W=0.1$.  This
choice of bin sizes results in a $24\times 15$ grid containing $360$
bins $dMdW$, of which 144 contain galaxies. The mean occupancy of the
bins is therefore 7 galaxies per bin.
 In Figure~\ref{2d.fig} we
also plot the data points on which the analysis is based.  It is clear
that the low mass bins contain very few galaxies, which explains the
noisy character of the lower left part of the figure.  Not surprisingly,
there is a strong correlation between the \hi\ mass of a galaxy and its
observed velocity width.  Note that the plotted velocity width is the
observed width of the \hi\ signals, uncorrected for galaxy inclination.

\vbox{
\vspace{0.1cm}
\begin{center}
\leavevmode
\hbox{%
\epsfxsize=\hsize
\epsfbox[19 143 569 695]{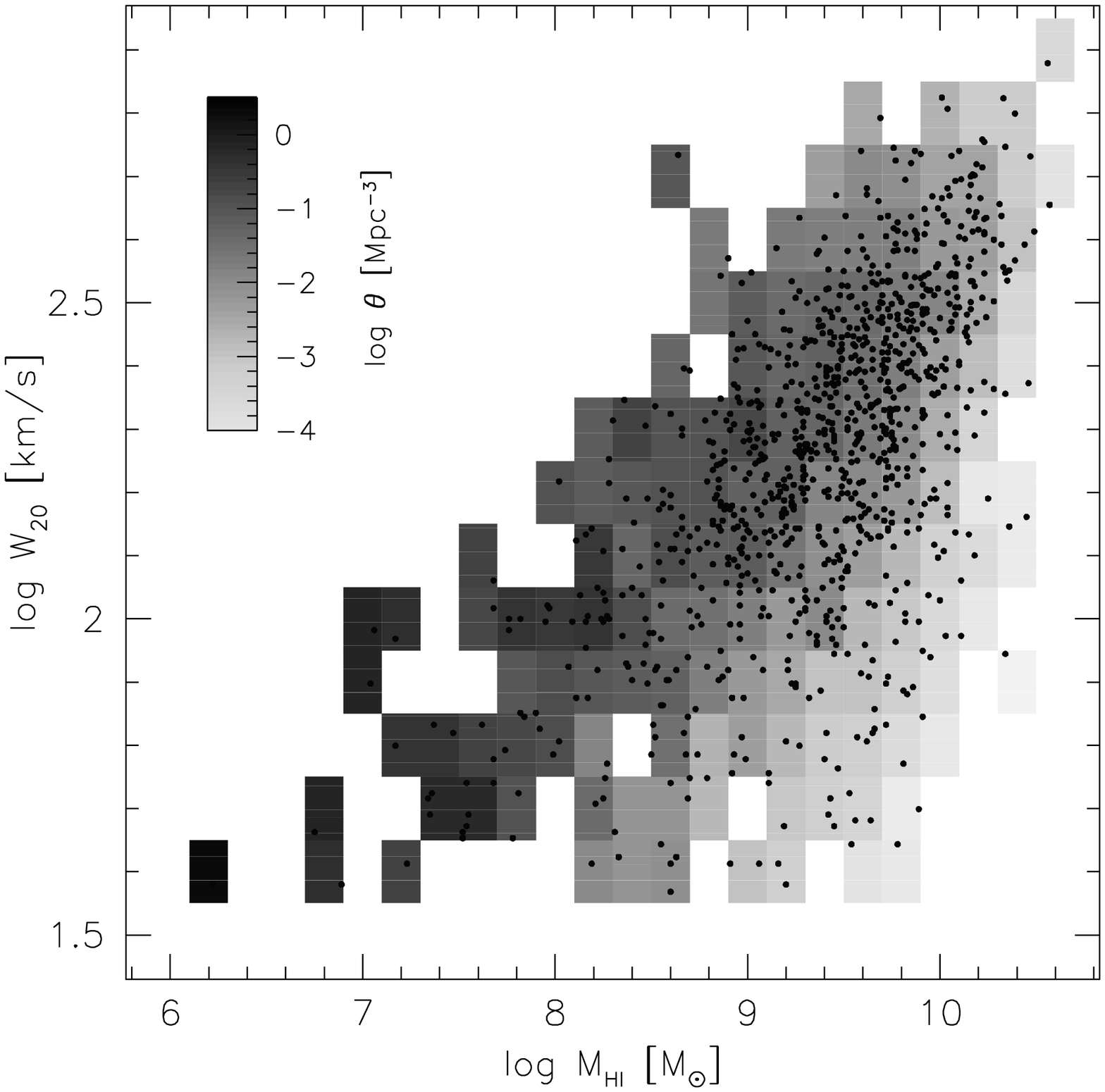}}
\figcaption{Bivariate galaxy density distribution in the ($\mhi,w_{20}$)
plane,
derived via the 2DSWML method. The greyscales are on a logarithmic
scale and represent the space density per decade of \mhi\ and decade of
$w_{20}$.  
The distribution of data points is shown by
black dots. 
\label{2d.fig}}
\end{center}}

A comparison between the greyscales and the black points illustrates
that a peak flux limited sample is, at a particular \hi\ mass, slightly
biased toward galaxies with low velocity widths, or alternatively, at a
particular velocity width, slightly biased toward galaxies with high
\hi\ masses.  Although the survey contains a large fraction of high \hi\
mass, low velocity width galaxies, the true space density of these
objects is very low. 

The \hi\ mass function can now be found by applying Eq.~\ref{theta_m.eq}
to the 2-dimensional density distribution $\theta_{jk}$.  The result is
shown in Figure~\ref{himfswml2d.fig}, where the dots show the measured
space density of objects per decade of \hi\ mass and the solid line is
the best fit Schechter function with parameters
 $\alpha=-1.30 \pm 0.08$,
 $\log (\mhis/\msol)=9.79 \pm 0.06$, and 
 $\theta^*=(8.6 \pm 2.1)\times 10^{-3} \,{\rm Mpc}^{-3}$.
  The Schechter function provides an excellent fit to the data.  In fact,
thanks to the large sample size and hence small Poisson errors, we
convincingly show here for the first time that the \hi\ mass function of
galaxies can be satisfactorily described by a Schechter function.
Note, however, that the shape and the normalization of the \himf\ are
determined without using a Schechter function as an assumption about the
intrinsic shape of the \himf.

\vbox{
\vspace{-0.0cm}
\begin{center}
\leavevmode
\hbox{%
\epsfxsize=\hsize
\epsfbox[32 250 566 692]{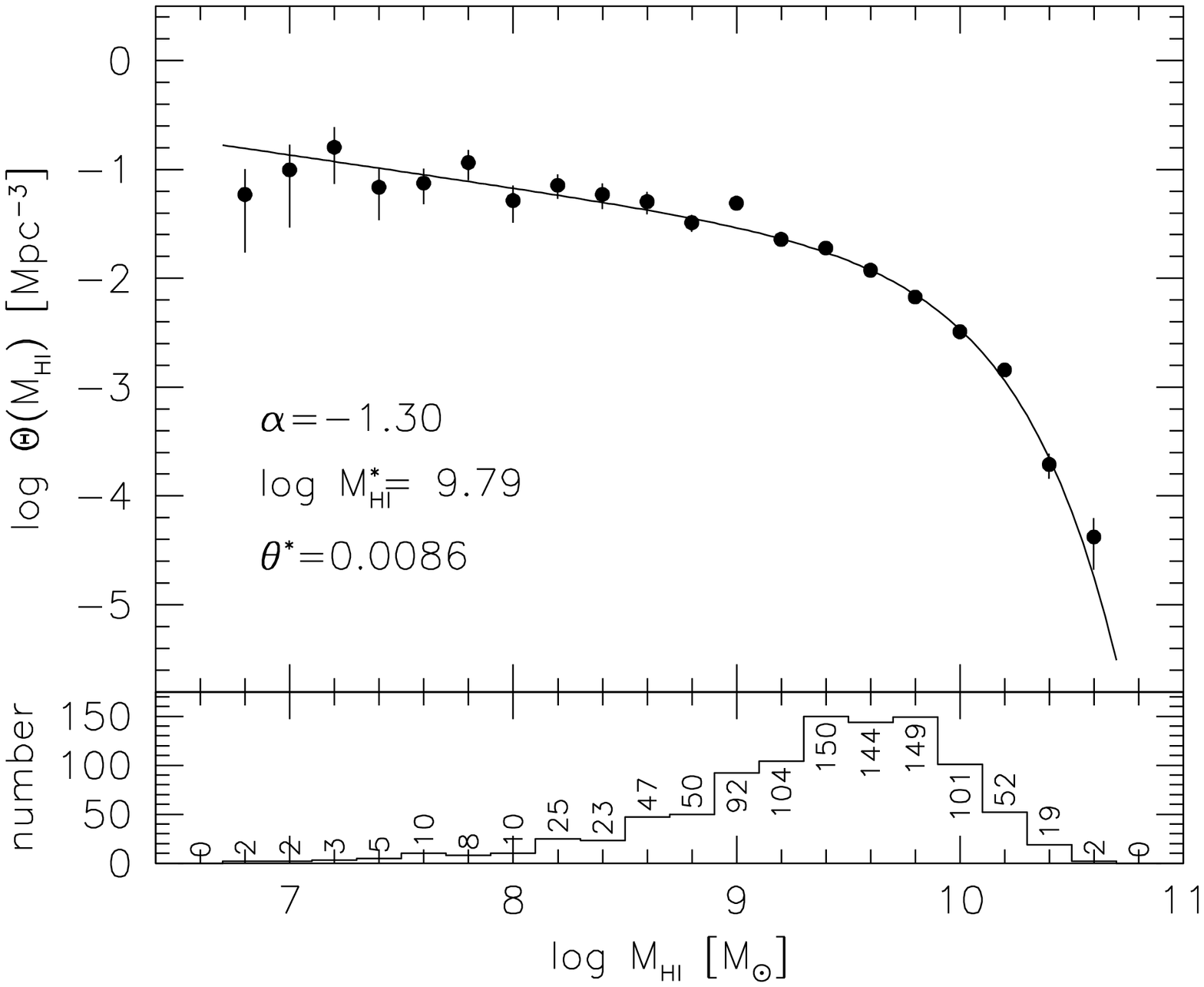}}
\figcaption{The BGC \hi\ mass function.
 {\em Bottom panel:\/} Distribution of \hi\ masses in the BGC per
bin of 0.2 dex.
{\em Top panel:\/}
Points and errorbars show the space density of objects derived with 
the 2DSWML method. The solid
line is a best-fit Schechter function which has been determined by $\chi^2$ 
minimalization. The best fit Schechter parameters are also
shown. \label{himfswml2d.fig}}
\end{center}}

The uncertainties on the Schechter parameters are determined with the
jackknife error estimator (Lupton 1993).  This is simply done by
dividing the sample into 24 equal regions of sky and calculating the
\hi\ mass function 24 times, leaving out a different region each time. 
A Schechter function is fitted to each mass function and the errors on
the parameters are found by
 \begin{equation}
 \sigma_x^2=\frac{N-1}{N}\Sigma_i(x-\bar{x})^2, \end{equation} 
 where $N=24$.  The quoted uncertainties therefore include statistical
errors and incorporate the effects of large scale structure, but they do
not include systematic errors.

As is normally the case in these fits, the errors in $\alpha$ and $\log
\mhi$ are correlated in the sense that steeper faint-end slopes imply
higher values of $\log \mhi$.  The correlation coefficient between
$\alpha$ and $\log \mhi$ is $r=-0.28$, but due to the small number of
jackknife samples ($N=24$), the error on $r$ is large. 

The 2-dimensional distribution function $\theta_{jk}$ can also be
integrated along lines of constant velocity width to obtain the
`velocity width function', the space density of objects as a function of
velocity width, or, after inclination correction, as a function of
rotational velocity.  This analysis will be the topic of a forthcoming
paper. 

\subsection{Biases in the calculation of the \himf}
 In the following we investigate the influence of various selection effects
on the shape of the \himf\ and on the measurement of the integral \hi\
mass density. The effects of the biases are summarized in Table~\ref{bias.tab}.

\subsubsection{Noise on \hi\ detection spectra}
 The BGC consist of the brightest galaxies in the total HIPASS galaxy
sample and therefore the signal to noise of the detections is high.  The
average rms noise in the HIPASS data is 13 mJy/beam per 18.0~km/s, which
means that with the BGC detection limit of 116 mJy the lowest possible
S/N level would be 9.  In reality, because the detections are spread out
over more than one resolution element, the S/N ratios are generally
higher.  However, since most detections are much wider than one
resolution element, the measured peak flux is often an overestimation of
the galaxy's noise-free peak flux.  This becomes increasingly important
for flat-topped profiles in which a large fraction of the channels are
close to the peak flux. 

We test this selection bias by adding Gaussian noise with a rms
dispersion of $13$ mJy to the \hi\ profiles of synthetic galaxy
samples which resemble the BGC, and selecting galaxies by their measured
peak flux.  Next we calculate \himfs\ from these samples using the
2DSWML method.  Two extreme situations are tested, one in which all
profiles are Gaussian, and one in which all profiles are double-horned. 
Figure~\ref{noise.fig} shows the true \himf\ of the simulated galaxies
as a solid line, the recovered \himf\ for double-horned profiles as open
circles and the recovered \himf\ for Gaussian profiles as crosses. 

\vbox{
\vspace{-0.1cm}
\begin{center}
\leavevmode
\hbox{%
\epsfxsize=\hsize
\epsfbox[32 325 566 692]{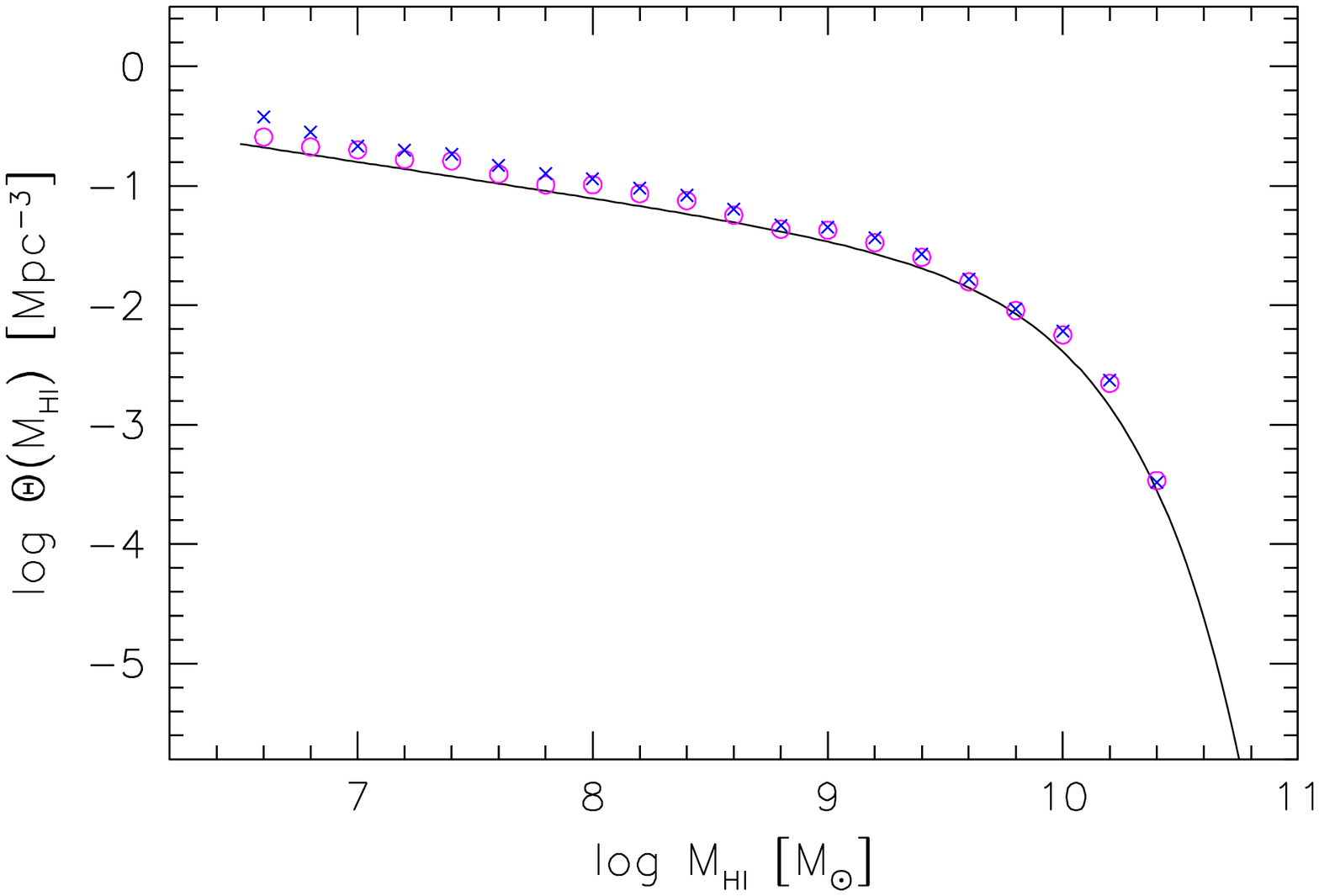}}
\figcaption{The effect of noise on the detection spectra on the \himf\
determination.  The solid line shows the input \himf\ with a low mass
slope of $\alpha=-1.3$ that has been used to create synthetic galaxy
samples.  Noise with an rms fluctuation of 13 mJy has been added to the
synthetic galaxy spectra.  The points show the recovered 2DSWML \himf\
assuming Gaussian profiles (crosses) and double-horned profiles
(circles).
\label{noise.fig}}
 \end{center}}

For both cases the selection bias causes a small global overestimation
of the \himf.  For the Gaussian profiles, $\alpha$ is overestimated by
$\Delta\alpha \approx 0.03$, whereas for the double-horned profiles the
change in $\alpha$ is negligible.  For the BGC, which is dominated by
galaxies showing double-horned profiles, the integral \hi\ mass density
$\rho_{\rm HI}$ is probably overestimated by less than $15\%$. 

\subsubsection{The Eddington effect}
 Another possible source of error in the \hi\ mass function is the
so-called `Eddington effect'.  This effect causes a steepening of the
low mass slope due to distance uncertainties.  For the vast majority of
the BGC galaxies distances are calculated from their recessional
velocities and peculiar velocities can introduce errors in these
distance estimates and hence in the \hi\ masses.  In a relatively nearby
sample such as the BGC the Eddington effect is potentially
much more important than for deeper surveys like the AHISS survey,
for which the peak of the galaxy distance distribution is much higher
and relative distance uncertainties are smaller.

To test the severeness of the Eddington effect, we add Gaussian noise to
the recessional velocities in our synthetic galaxy samples and select
galaxies from the samples similar to the selection of the BGC.  The
results are summarized in Figure~\ref{eddington.fig}.  The solid line
shows the input \himf\ and the symbols show the recovered \himfs\ for
samples for which dispersions of 50 (squares), 100 (triangles), 200
(crosses) and 400 (circles) \kms\ are added.  Each simulation is
performed 100 times.  The 2DSWML method has been used to calculate the
\himfs. 

Figure~\ref{eddington.fig} shows that the Eddington effect is a
potentially important bias in our analysis.  If a velocity dispersion of
200~\kms\ is added to our synthetic galaxy samples, the measured \himf\
slope increases from $\alpha=-1.30$ to $\alpha=-1.43$.  For a velocity
dispersion of 100~\kms\, $\alpha$ rises to $-1.36$, for 50~\kms\ the
change in $\alpha$ is no longer measurable.  As expected, the Eddington
effect has almost no influence on the measured space density of sources
around $\mhi=\mhis$, therefore the effect on the integral \hi\ mass
density $\rho_{\rm HI}$ is small.  The inset in
Figure~\ref{eddington.fig} shows the recovered selection function $S(D)$
for each of the simulations, compared to the selection function of a
galaxy sample without added noise to the distances (dashed line).  The
selection functions have been multiplied with $D^2$ and are normalized
to a peak value equal to unity.  
The x-axis is logarithmic to better show the
effects on the selection function at small distances.  This clearly
illustrates that the Eddington effect artificially steepens the
selection function at small distances. 

\vbox{
\vspace{0.2cm}
\begin{center}
\leavevmode
\hbox{%
\epsfxsize=\hsize
\epsfbox[32 324 566 692]{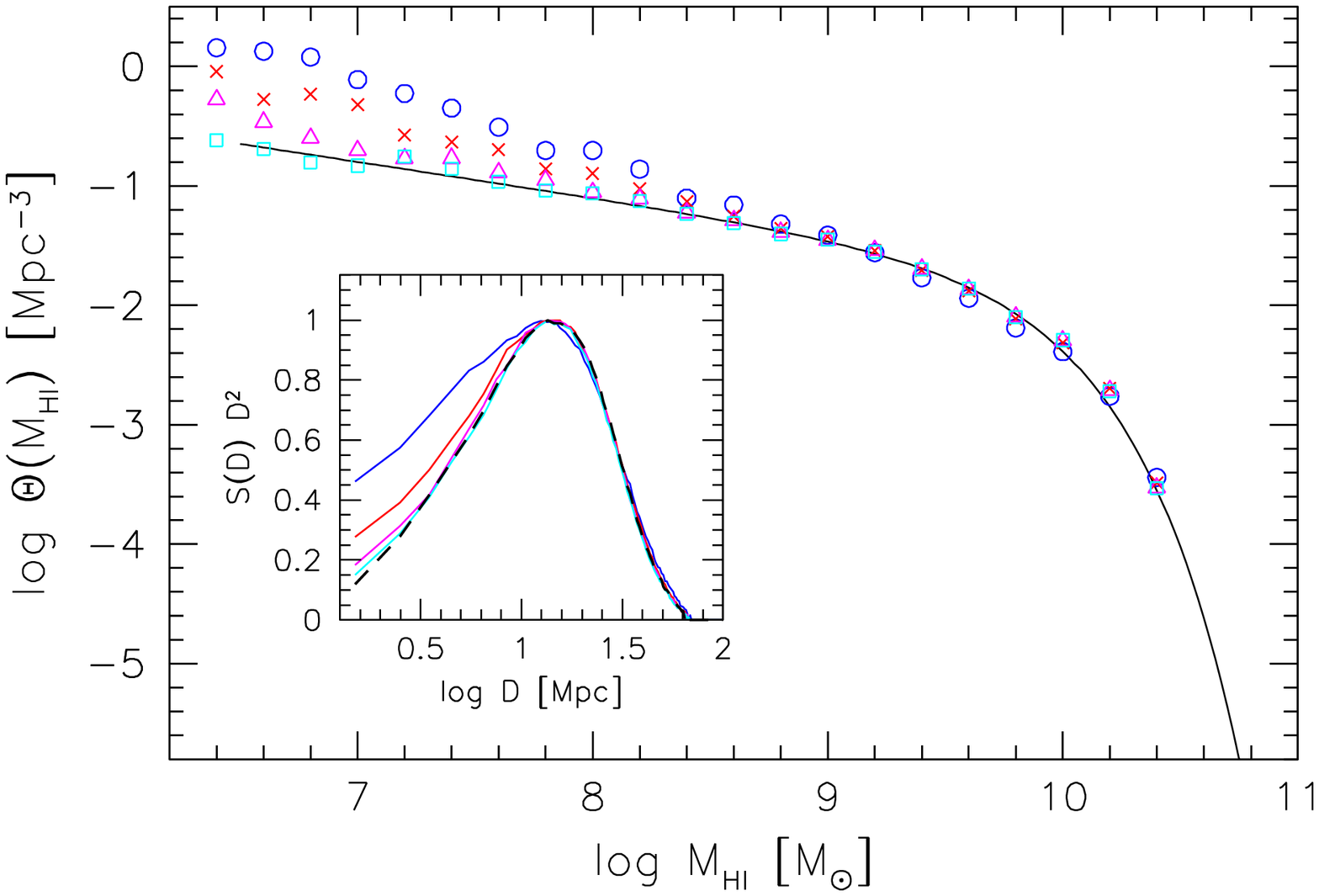}}
\figcaption{The Eddington effect.  The solid line shows the \himf\ of
simulated data sets, with a low mass slope of $\alpha=-1.3$.  The
symbols show the recovered 2DSWML \himfs\ after Gaussian noise is added to
the recessional velocities of the galaxies.  Squares are for a
dispersion of 50~\kms, triangles for 100~\kms, crosses for 200~\kms, and circles for 400~\kms. 
\label{eddington.fig}}
 \end{center}}

Our model of adding Gaussian noise to the recessional velocities
of galaxies oversimplifies the true peculiar velocity distribution.
The real dispersion is a function of local density (Strauss, Ostriker \&
Cen 1998) and should be directional to some degree, instead of random.
However, Sheth \& Diaferio (2001) recently showed that the expected
peculiar velocity distribution in CDM simulations is nearly Gaussian.
 The 1-D pairwise peculiar velocity dispersion $\sigma_{12}$ is found to
be $\approx 300~\kms$ on scales of $1 h_{100}^{-1}$ Mpc (e.g.  Jing,
B{\"o}rner \& Suto 2002) and lower on smaller scales.  This value
compares to a velocity noise of $\sigma=2^{-1/2} \sigma_{12}=212~\kms$. 
Clusters of galaxies contribute significantly to this high value and
the velocity field outside of clusters, where late-type galaxies
dominate the statistics, is known to be much colder (e.g.  Strauss et
al.  1998).  For example, Willick et al.  (1997) find $\sigma=125~\kms$
for local ($cz < 3000~\kms$) spiral galaxies.  More locally, the
velocity field is even colder than this: the peculiar velocity
dispersion of galaxies within 5 Mpc of the Milky Way Galaxy is only
60~\kms\ (Schlegel et al.  1994).  Considering the latter two values, our
simulations with dispersions of $50\, \kms$ and $100\, \kms$ probably define
the boundaries of realistic test of the Eddington effect.  From
Figure~\ref{eddington.fig} we conclude that our estimated \himf\
probably overestimates the steepness of the faint end slope $\alpha$
with $\Delta\alpha < 0.05$.  Over the range of \hi\ masses where we can
reliably measure the \himf, the \hi\ mass density $\rho_{\rm HI}$ is
maximally overestimated by 8\%. 

Our results are different from Rosenberg \& Schneider (2002) who find
that for their sample the Eddington effect is unimportant even for a
dispersion of 600 \kms.  This difference can be explained by the
difference in survey depth between the BGC and the ADBS.  In the
analysis of the deep HIPASS catalog, the Eddington effect will be
less important. 

Of course, the \hi\ mass estimates of the nearest galaxies suffer the
most from the distance uncertainties.  A solution would be to impose a
lower distance limit to the galaxy sample above which the distance
uncertainties are believed to be small.  However, for samples other than
integrated flux limited, imposing a lower distance limit causes a slight
drop of the \himf\ at low \hi\ masses.  This happens because in the low
\hi\ mass bins only those galaxies will be selected that have a high
peak flux compared to other galaxies of the same \hi\ mass at the same
distance. 

\subsubsection{Inclination effects} \label{mininc.sec}
 The inclination of galaxies might lead to two independent biases in the
\himf\ determination.  One effect, \hi\ self-absorption, is discussed in
the next subsection (\ref{selfab.sec}).  Here we investigate the effect
that in the galaxy selection procedure a minimal velocity width $w_{\rm
min}$ is applied, below which galaxy signals can not be reliably
distinguished from radio frequency interference (RFI).  Since galaxies
with low inclinations have smaller velocity widths, this selection
effect might lead to a underrepresentation of face-on galaxies.  Via the
Tully-Fisher relation (Tully \& Fisher 1977) and the relation between
\hi\ mass and optical luminosity (e.g., Roberts \& Haynes 1994), 
it is known that galaxies with
low \hi\ masses have lower velocity widths.  The selection bias
therefore becomes progressively more important for dwarf galaxies, possibly
causing a flattening of the \hi\ mass function.  Lang et al.  (2002)
were the first to calculate the severeness of this effect and concluded
that for their sample of galaxies the \himf\ could be
underestimated by 18.7\% at $\mhi=2\times 10^7\,\msol$.  
Here we make specific calculations for the BGC. 

The velocity width measured at 50\% of the peak flux, $w_{50}$, can be
written as 
 \begin{equation} \label{w50.eq}
 w_{50}=(w_0^2 \sin^2 i+w_{\rm tur}^2)^{1/2} + w_{\rm inst},
 \end{equation}
where $w_0$ is the intrinsic velocity spread due to rotation, $w_{\rm tur}$ is
the velocity width due to turbulence in the gas layer, and $w_{\rm inst}$
is the contribution of instrumental broadening to the velocity width. 
For  $w_{\rm tur}$ we adopt the conservative value of
6 \kms\ from Verheijen \& Sancisi (2001) and for $w_{\rm inst}$ we use the
Bottinelli et al.  (1990) estimate of $w_{\rm inst}=0.13\times \delta v$, where
$\delta v$ is the velocity resolution. For our case, where $\delta v=18~\kms$, 
we find that $w_{\rm inst}=2.3~\kms$. 

For detections to be included in the BGC, signal must be found in at
least two consecutive channels.  The BGC is limited to sources with
$S_{\rm p} >116~ \rm mJy$, but the original search limit was
approximately 50\% lower, at 60 mJy (see Koribalski et al.  2003).  We
therefore assume that the minimal velocity width for inclusion in the
BGC, $w_{\rm min}$, applies to the 50\% level and hence $w_{50}>w_{\rm
min}=26.4~\kms$. 
Following Lang et al.  (2002), we can
calculate from Eq.~\ref{w50.eq} the minimal inclination angle $i_{\rm
min}$ that a galaxy needs to have for it to be included in the sample:
 \begin{equation}
 i_{\rm min}={\rm arcsin} \{ \frac{(w_{\rm min}-w_{\rm inst})^2-w_{\rm tur}^2}{w_{0}^2} \}^{1/2}.
 \end{equation}
Using the values given above we find that $i_{\rm min}=13.5^\circ$ for
$w_0=100~\kms$ and $i_{\rm min}=6.7^\circ$ for $w_0=200~\kms$. Recall
that $w_0$ is the total velocity width of a galaxy, which is twice the
rotational velocity.

From this we can estimate $\zeta$,
the fraction of galaxies potentially missed at every $w_0$,
by integrating from $i=0$ to $i=i_{\rm min}$ over a randomly oriented
sample:
 \begin{eqnarray}
 \zeta= \int_0^{i_{\rm min}} \sin i\, di=
 1-\cos i_{\rm min}=\nonumber\\
1-\{1- \frac{(w_{\rm min}-w_{\rm inst})^2-w_{\rm tur}^2}{w_{0}^2} \}^{1/2}.
 \end{eqnarray}
We find that the values for $\zeta$ are very small: $\zeta= 2.7\%$
($0.7\%$) for $w_0=100~\kms$ (200~\kms). We note that our adopted
values for $w_{\rm inst}$ and $w_{\rm tur}$ are very conservative, and much
higher values can be found in the literature. With higher values of 
$w_{\rm inst}$ and $w_{\rm tur}$ the resulting $\zeta$ would become even smaller.

\vbox{
\vspace{-0.0cm}
\begin{center}
\leavevmode
\hbox{%
\epsfxsize=\hsize
\epsfbox[32 325 566 692]{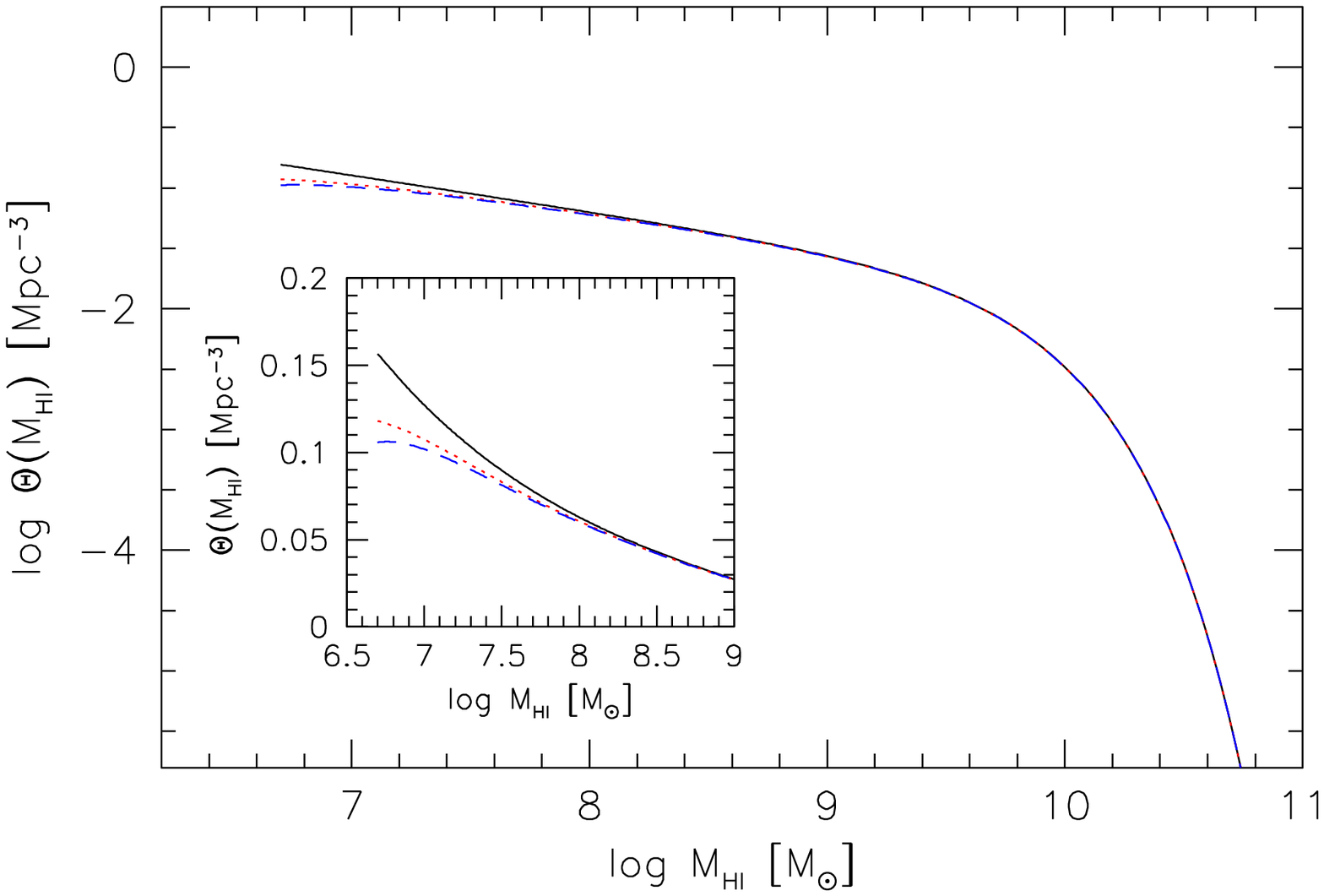}}
\figcaption{The effect of minimal velocity width on the \himf\ measurement. 
The input \himf\ is drawn as a solid line.  The 2DSWML \himf\ that is measured
if face-on galaxies are missed due to the minimal velocity width
requirement, is shown as a dotted line.  The dashed line is the same, but
for $w_{\rm tur}=w_{\rm inst}=0$.  The inset shows $\Theta(\mhi)$ on a linear
scale to better show the differences between the lines. 
 \label{minincl.fig}}
 \end{center}}

Finally, in order to express $\zeta$ as a function of \hi\ mass, we
need to adopt a relation between $\mhi$ and $w_0$.  Lang et al.
(2002) adopt the relation $w_0=0.42 \mhi^{0.3}$, where in their case
$w_0$ is derived from $w_{20}$.  Here we use $w_{50}$ and find that
$w_0=0.35 \mhi^{0.3}$ is a better fit to the BGC data.  We note,
however, that the scatter in the correlation between $\mhi$ and $w_0$
is very large. An expression of $\zeta$ as a function of \mhi\ should
therefore be regarded as illustrative. If we boldly apply the
relation, we find that $\zeta=10\%$ $(5\%)$ for $\mhi=2\times
10^7~\msol$ $(5\times 10^7~\msol)$.

In Figure~\ref{minincl.fig} we show the effect of the inclination bias. 
The solid line is an \himf\ with $\alpha=-1.30$ and $\log \mhis=9.80$,
and the dotted line is what would have been measured if galaxies with
low inclination angles were missed in the survey.  The dashed line is
the same, but here we test the extreme case that 
$w_{\rm tur}$ and $w_{\rm inst}$ are negligible.
The inset shows
$\Theta(\mhi)$ on a linear scale to better show the differences between
the lines.  The inclination bias only becomes apparent at very low \hi\
masses, implying that only the last two bins of our measured \himf\ are
slightly affected by the bias.  Over the \hi\ mass range where the
\himf\ is measured, the mass density $\rho_{\rm HI}$ is only
underestimated by approximately $1\%$. 

\subsubsection{\hi\ self-absorption} \label{selfab.sec}
 In the Milky Way Galaxy the cold neutral phase of the interstellar
medium can be seen as \hi\ self-absorption (HISA) against the \hi\
background (e.g., Knapp 1974, Gibson et al. 2000).  It is not well
known to what extent this HISA reduces the 21cm emission line flux of
external galaxies.  Deep, high resolution 21cm maps of nearby spirals
(Braun 1997) show that between 60 and 90\% of the \hi\ flux comes from
of a network of cool gas ($T<300 \,\rm K$), which is expected to have a
high opacity. However, not may instances of \hi\ absorption against
background continuum sources is observed in these regions. Dickey et
al. (2000) find that in the Small Magellanic Cloud the abundance of
cool-phase gas is only 15\%, a factor of two lower than what is found
in the Milky Way Galaxy.  The temperature of this gas is very low,
typically 40~K or less. From detailed absorption studies they
calculate the HISA correction as a function of column density, and find
that the correction is negligible for $\nhi<3\times 10^{21}~\rm
cm^{-2}$, and rises to 1.4 at $\nhi=10^{22}~\rm cm^{-2}$. These low
correction values agree with the analysis of Zwaan, Verheijen and
Briggs (1999), who find that the column density distribution function
of galaxies follows $\nhi^{-3}$, expected for optically thin gas, up to
$\nhi=8\times 10^{21}~\rm cm^{-2}$.

Haynes \& Giovanelli (1984) plotted the \hi\ mass
of galaxies as a function of inclination and measured $f_{\rm HI}$,
the ratio of HISA corrected to measured flux, for different
morphological types.  Zwaan et al.  (1997) considered the effect of
HISA on the \himf, and based on Haynes \& Giovanelli's (1984) result
they found that the average value of $f_{\rm HI}$ for a randomly
oriented galaxy sample is approximately 1.10.  They concluded that
$\mhis$ and $\thetas$ would increase by no more than 10\% if HISA
effects were taken into account.  The integral density $\rho_{\rm HI}$
could be underestimated by $\sim 15\%$, maximally.  Lang et al.
(2002) addressed the issue of HISA by plotting the distribution of
inclination angles of their HIJASS galaxies, and found that the number
of highly inclined galaxies is lower than what is expected for a
randomly oriented sample.  From this they derive that $f_{\rm HI}$ is
1.25 and that $\rho_{\rm HI}$ could be underestimated by as much as
$25\%$.

In Figure~\ref{incdist.fig} the distribution of
the cosine of inclination angle $i$ of cataloged BGC galaxies is drawn
as a solid line (see Jerjen et al. 2003 for optical properties of the
BGC sample).  
For a randomly oriented sample the distribution should
be flat.  The BGC distribution is clearly under-abundant at low and high
inclination, suggesting the effects of HISA at high $i$ and the minimal
velocity width effect (see Section~\ref{mininc.sec}) at low $i$. 
However, also shown in Figure~\ref{incdist.fig} is the $\cos i$
distribution for the 1000 brightest southern optically selected galaxies
from LEDA (dashed line), and the 1000 brightest southern optically selected
galaxies with available 21cm data (dotted line).  All distributions are
indistinguishable.  This implies that the distribution of $\cos i$ is
not determined by HISA and minimal velocity width, but by the
inclination measurements available from the literature.  As was
discussed by Huizinga \& van Albada (1992), purely circular isophotes
are seldom observed in galaxies, leading to an apparent
underrepresentation of face-on disks.  On the other end of the
distribution, three different effect cause a deficit of galaxies. 
Firstly, due to dust obscuration, highly inclined galaxies drop out of the
optical sample.  These galaxies are also likely to be missed when our
\hi\ selected sample is cross-correlated with optical catalogs.
Secondly, the intrinsic thickness of galaxies causes uncertainties in $i$,
especially for highly inclined galaxies. This effect causes the spike at
$\cos i=0$ and may cause a deficit at low $i$. Thirdly, HISA might
cause a underrepresentation of high $i$ galaxies. However, since the
optical information on the BGC is incomplete, it is not possible to
disentangle the effects of HISA and optical dust extinction.
We therefore conclude that based on the information at hand, no
meaningful measurement of the effects of HISA can be made.
In absence of reliable intrinsic measurements, we
adopt the externally measured value of 15\% underestimation of
$\Omega_{\rm HI}$, derived from the results of Zwaan et al.  (1997). 

\vbox{
\vspace{-0.0cm}
\begin{center}
\leavevmode
\hbox{%
\epsfxsize=\hsize
\epsfbox[32 327 566 692]{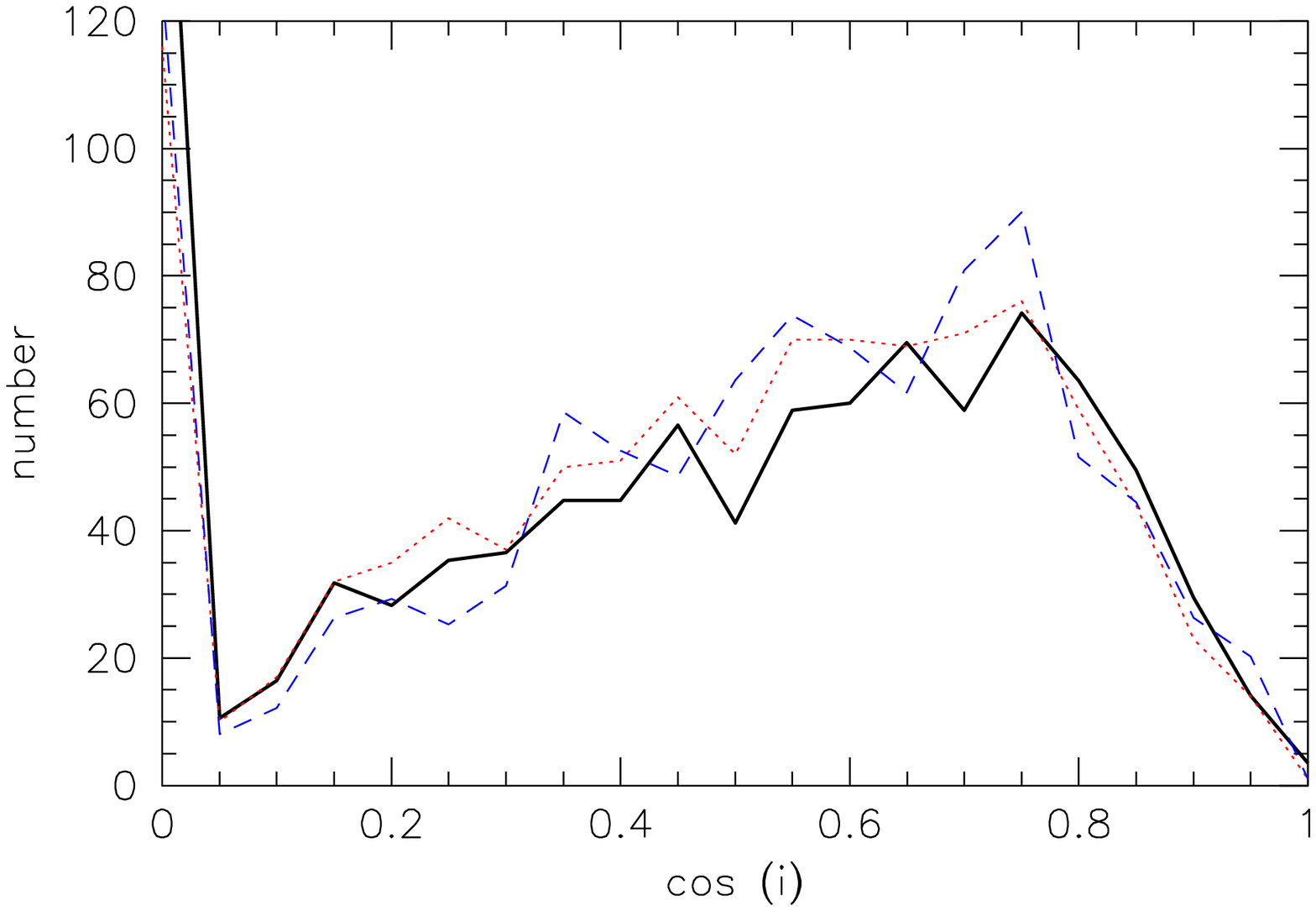}}
\figcaption{Distribution of cos inclination angle.  The solid line is the
distribution for BGC galaxies with optical information.  The dashed line
is for the 1000 brightest southern optically selected galaxies from
LEDA, and the dotted line is for the 1000 brightest southern optically
selected galaxies with available 21cm data.
\label{incdist.fig}}
\end{center}}

\subsubsection{Confusion} 
The relatively coarse spatial resolution of the HIPASS survey
($15\farcm5$) causes confusion of some signals with neighboring galaxies.
Via ATCA follow-up observations and comparison with literature values,
Koribalski et al.  (2003) identified in the BGC 11 compact groups and
44 pairs of galaxies whose signals might chance to coincide in
frequency. Furthermore, there are 67 sources that are flagged as
`confused'.

 It is expected that some of the confused sources would not make it into
the BGC if they were resolved because their peak flux would drop below
the selection limit.  In that case we could overestimate the normalization of
the \himf\ as artificial sources would enter the sample.  On
the other hand, if the unconfused sources had peak fluxes above the
selection limit, galaxies would shift to higher mass bins, leading to an
overestimation of \mhis. 

We test the two extremes by splitting all sources marked as `confused'
or `pair' into two sources of equal \hi\ mass, and distinguish two cases: 
 1) the peak flux of the new sources is equal to that of their parent
source, but their velocity width is halved,
 2) the velocity width of the new sources is equal to that of their
parent source, but the peak flux is halved. In this case only the new
sources that have $S_{\rm p}>116~ \rm mJy$ are counted.
 For case 1) we find that \mhis\ drops by $\sim 15\%$, \thetas\ increases
by $\sim 15\%$, and $\alpha$ does not change significantly.  Since the
changes in $\thetas$ and \mhis\ are balanced, there is no nett effect on
the \hi\ mass density $\rho_{\rm HI}$.  In case 2) the changes in
$\thetas$ and \mhis\ are comparable, but additionally there is a
marginal decrease in $\alpha$ of $\approx 0.02$.  As in this case some
sources drop below the detection limit, $\rho_{\rm HI}$ drops slightly
with $\approx 2\%$. 

\subsubsection{Cosmic variance}
 In this section we investigate to what extent the shape of the \himf\
depends on the region of the sky that is investigated.  Previous blind
\hi\ surveys were based on relatively small regions of sky (66
degrees$^2$ for the AHISS and 430 degrees$^2$ for the ADBS), whereas the
present analysis is based on a survey covering $2\times10^5$
degrees$^2$.  This large area would be expected to guarantee a fair
sampling of the local volume, but on the other hand, the BGC is a
shallow survey compared to the AHISS and the ADBS. 

Figure~\ref{cosmic_var.fig} shows 2DSWML \himfs\ for the four different
quadrants of the southern sky.  Around $\mhi=\mhis$, the variation in
the \himfs\ is only mild, but at $10^8~\msol$ the estimated space density
varies with a factor 5, with the first quadrant being the most deviant
one.  Not only is the estimated average density lowest there, but the
low mass end is also flatter.  The cosmic variance is also reflected in
the number of sources in each quadrant.  The third quadrant contains 313
sources, whereas the fourth quadrant contains only 159 sources. 

\vbox{
\vspace{-0.1cm}
\begin{center}
\leavevmode
\hbox{%
\epsfxsize=\hsize
\epsfbox[32 320 566 692]{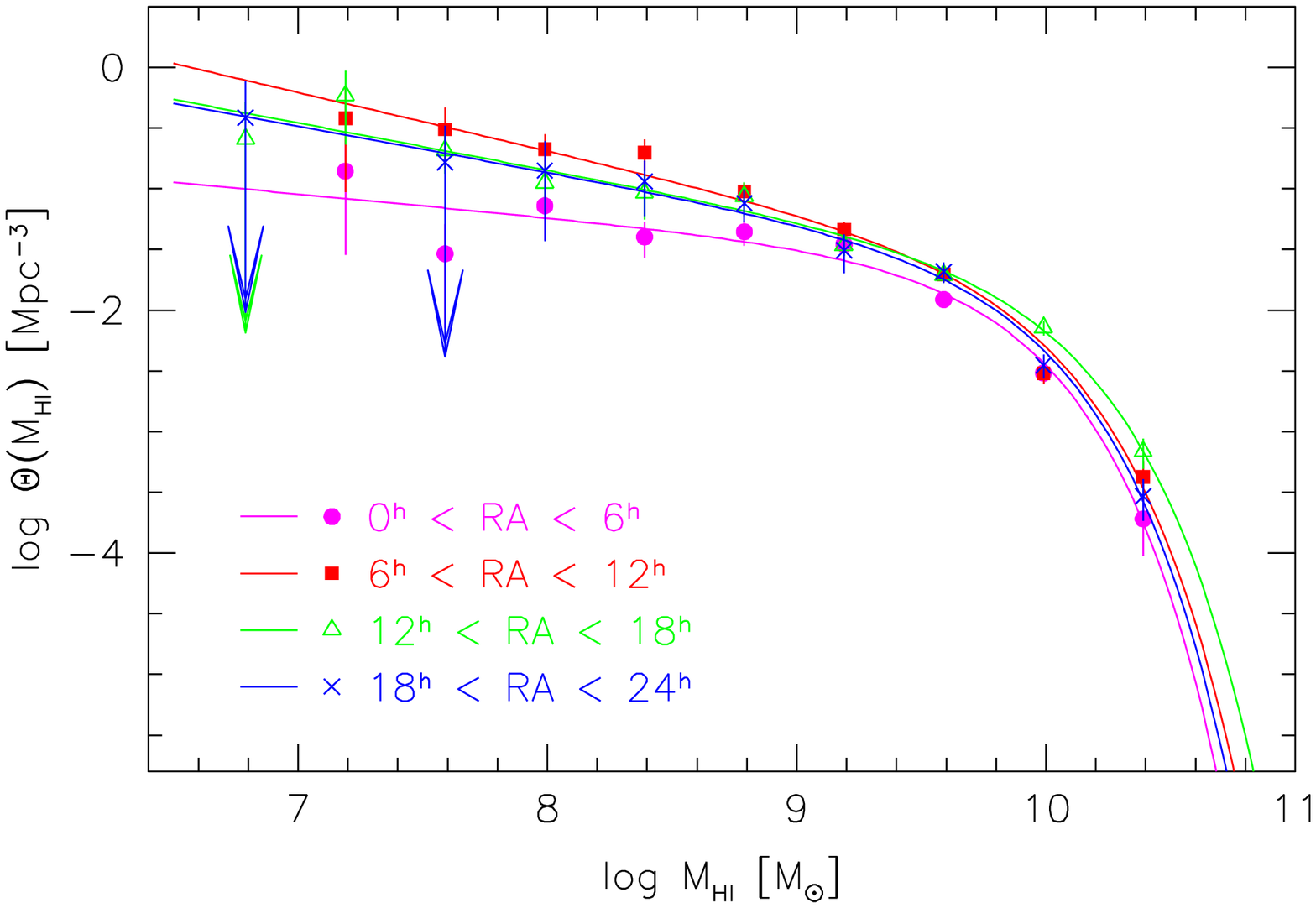}}
\figcaption{The effect of cosmic variance on the \himf\ shape.
The different symbols correspond to \himf\ determinations in four
different quadrants of the southern sky. 
\label{cosmic_var.fig}}
\end{center}}

This exercise shows that selecting small regions of sky for galaxy
surveys can introduce substantial uncertainties in the estimated space
density of galaxies.  We stress, however, that several properties of a
survey define its immunity against large scale structure: the
sensitivity, the covered area of sky, and the shape of the covered
area.  The former two properties together define the search volume of
the survey.  Surveys like the AHISS and the ADBS cover small
declination ranges, but $24^{\rm h}$ in RA, whereas the HIPASS SCC
survey covers more square degrees but is concentrated only on the
South celestial cap. In addition, the immunity against large scale
structure depends on whether the search volume extends well beyond the
local overdensity surrounding the Milky Way Galaxy.

\begin{deluxetable*}{lll}
\tabletypesize{\scriptsize}
\tablecaption{Biases in the HIMF determination \label{bias.tab}}
\tablewidth{0pt}
\tablehead{
\colhead{Bias} &  
\colhead{Effect on \himf} &  
\colhead{Effect on $\Omega_{\rm HI}$}   
}  
\startdata
Selection bias 		& overall increase ($\approx 10\%$) & $\lesssim 15\%$ over \\
Eddington effect 	& steepening ($\Delta\alpha\approx- 0.05$)& $8\%$ over \\
\hi\ self-absorption 	& overall decrease ($\approx -10\%$) & $\lesssim 15\%$ under \\
Minimal velocity width  & flattening ($\Delta\alpha\approx 0.03$) & $1\%$ under \\
Confusion 		& increase in \mhis\ ($\approx 15\%$) and
  decrease in \thetas\ ($\approx 15\%$)     & none  \\
Cosmic variance 	& small error in shape 	& uncertainty $\sim 10\%$\\
\enddata
\end{deluxetable*}

\begin{deluxetable*}{lccccccccccc}
\tablecolumns{12} 
\tabletypesize{\scriptsize}
\tablecaption{Number of cells and volumes probed by blind \hi\ surveys \label{cells.tab}}
\tablewidth{0pt}
\tablehead{
\colhead{Sample} & \multicolumn{3}{c}{$\mhi\sim 10^8 \,\msol$} &
\colhead{} &  
\multicolumn{3}{c}{$\mhi\sim 10^9 \,\msol$} &
\colhead{} &  
\multicolumn{3}{c}{$\mhi\sim 10^{10} \,\msol$} \\
\cline{2-4} \cline{6-8} \cline{10-12}\\ 
\colhead{} & 
\colhead{$10^3$}& \colhead{$20^3$} & \colhead{volume} &
\colhead{} &  
\colhead{$10^3$}& \colhead{$20^3$} & \colhead{volume} &
\colhead{} &  
\colhead{$10^3$}& \colhead{$20^3$} & \colhead{volume} 
}  
\startdata
HIPASS BGC 	& 2&1   & 5	 && 30&4   & 90   && 1500&190  & 1600  \\
AHISS	        & 60&15 & 0.35$\rm ^a$  && 200&50 & 1.6$\rm ^a$ && 200&50  & 1.6$\rm ^a$	   \\
Arecibo Slice   & 3&1   & 0.25   && 30&8   & 4.2  && 50&12     & 6.5	   \\
HIPASS SCC      & 2&1   & 0.6    && 10&2   & 10   && 200&20    & 180	   \\
HIPASS HIZSS	& 2&1   & 0.25   && 14&4   & 4.7  && 100&24    & 84	   \\
ADBS		& 20&5  & 0.5    && 120&20 & 8.6  && 700&100   & 51	   \\
\enddata
\tablecomments{For each \hi\ mass, the first column gives the number
     of probed $10^3\, \rm Mpc^3$ cells and the second column gives the
     number of probed $20^3\, \rm Mpc^3$ cells. The third column gives
     the approximate search volumes in 1000 Mpc$^3$. All search
     volumes are calculated using the quoted rms noise levels in the
     respective papers and assume optimal smoothing and a relation
     between velocity width and \hi\ mass as given in Zwaan et al. (1997).\\
     $\rm^a$ Not counting sidelobes}
\end{deluxetable*}

To quantify these statements and test the sensitivity to large scale
structure of blind \hi\ surveys, we compute for each survey the number
of $10^3\,\rm Mpc^{3}$ and $20^3\,\rm Mpc^{3}$ cells that are probed for
galaxies with different \hi\ masses.  The sizes of 10 and 20~Mpc are
chosen because they span the typical sizes of voids in the local universe
(Hoyle \& Vogeley 2002; Plionis \& Basilakos 2002).  For a survey to be
insensitive to large scale structure it seems reasonable to require that
it probes several of such cells.  Note that in the calculation of the
number of cells we do not require that all cells are fully probed, but
instead we calculate the number of cells from which each survey selects
its detections.  The results are given in Table~\ref{cells.tab}, which also 
shows the search volumes for different \hi\ masses for all surveys.

An important conclusion is that all surveys, except for the strip
surveys AHISS and ADBS, select their $\mhi=10^8~\msol$ sources from only
one $20^3\, \rm Mpc^3$ cell, and at most a few $10^3\, \rm Mpc^3$ cells. 
At $\mhi=10^9~\msol$ most surveys cover a large number of cells,
although the shallow, large scale samples selected from HIPASS still
probe fewer cells than the other surveys.  At $\mhi=10^{10}~\msol$ all
surveys are probably insensitive to large scale structure effects. 
Since these are the \hi\ masses that dominate the total \hi\ mass
density (see Section~\ref{dens.sec}), measurements thereof based on
HIPASS data are very secure.

Of course, this analysis may be too complementary to the strip
surveys. Although these surveys go through many more cells than the
large scale surveys, their sensitivity to large scale structure is not
reduced as much as suggested by Table~\ref{cells.tab}, because their
cells are not fully sampled. This can also be seen from the number of
galaxies that the different surveys detect at each \hi\ mass.
The ADBS detected seven galaxies with $\mhi<10^8~\msol$, whereas the
BGC has 38 galaxies in that mass range. Surely, the larger search
volume and higher number of galaxies in the BGC must reduce its
susceptibility to the effects of large scale structure.

\subsection{Comparison with other methods}
 Here we compare the results of the 2DSWML analysis of the BGC with those
from other, more conventional estimators.  Figure~\ref{comparemeth.fig} shows
the results of the standard $1/V_{\rm max}$ method represented by open
circles, and the standard SWML application to a integrated flux limited
subsample ($S_{\rm int}>25~ \rm Jy\, km\, s^{-1}$) 
is shown by grey triangles. 
For comparison, the 2DSWML solution is also reproduced in this figure
by a solid line.  The lower panel of Figure~\ref{comparemeth.fig} shows the
averaged $V/V_{\rm max}$ values for each bin in \hi\ mass.  The circles
and triangles represent the total BGC and the integrated
flux-limited subsample, respectively, and values of $V_{\rm max}$ are
calculated on the bases of their respective selection criteria.

The \himfs\ from the 2DSWML method and the standard $1/V_{\rm max}$ method
are in very good agreement.  This is particularly striking since the
latter method makes no corrections for the effects of large scale
structure which are obviously present in our data (see
Koribalski et al. 2003).  The mean value of $V/V_{\rm max}$ for the $1/V_{\rm
max}$ method is $0.503 \pm 0.009$, supporting the fact that the sample is
statistically complete and that the effects of large scale structure
average out over the whole sample.  Zwaan et al.  (1997) tested the
effects of density variations on their $1/V_{\rm max}$ determination of
the \himf\ and also concluded that their calculation was insensitive to
large scale structure.  For \hi\ masses $<10^8\,\msol$, the $1/V_{\rm
max}$ method finds higher space densities than the 2DSWML method.  This
is most likely the result of the fact that these galaxies are all drawn
from the very local universe ($<10$ Mpc, see Figure~\ref{mhi_D.fig}),
which we know to be overdense.  Also the Poisson errors on these points
are very high since they are based on small numbers of galaxies. 

\vbox{
\vspace{-0.0cm}
\begin{center}
\leavevmode
\hbox{%
\epsfxsize=\hsize
\epsfbox[32 250 566 692]{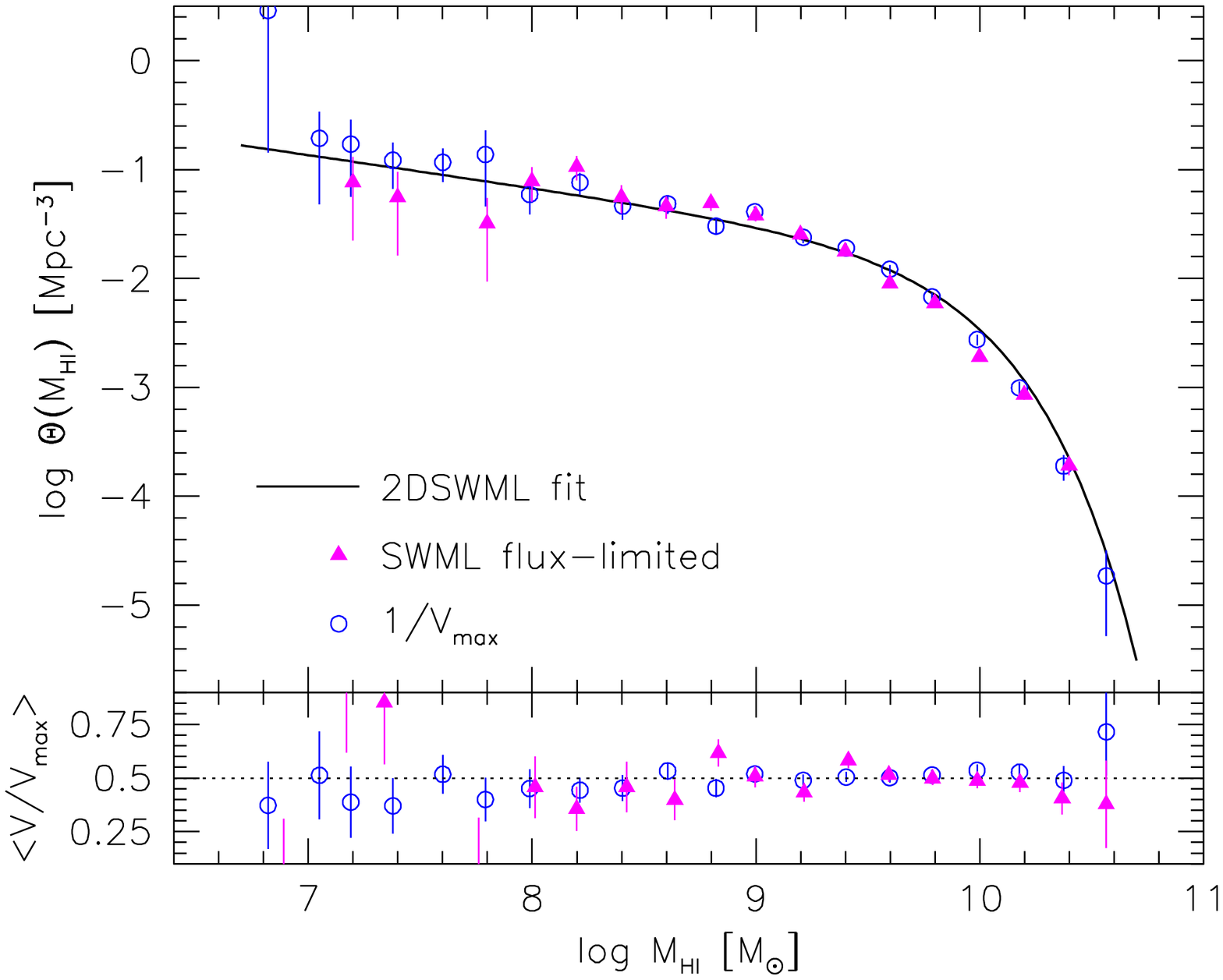}}
\figcaption{Comparison of different \himf\ estimators.  The
top panel shows the \himf\ of BGC galaxies derived via the $1/V_{\rm max}$
method as open circles and the standard SWML \himf\ of
an integrated flux limited subsample as grey triangles.  
The 2DSWML \himf\ from Figure~\ref{himfswml2d.fig} is reproduced as a
solid line.  The lower panel shows the median values of $V/V_{\rm
max}$ in each bin, with $1\sigma$ uncertainties.  Open circles again
represent the $1/V_{\rm max}$ method and grey triangles the standard
SWML method. \label{comparemeth.fig}}
 \end{center}}

The SWML \himf\ is also in good agreement with both other methods, but the
uncertainties on the space densities for $\mhi<10^8\,\msol$ are large
because the integrated flux limited subsample contains only a small
number of low mass galaxies.  The mean $V/V_{\rm max}$ value is $0.500
\pm 0.013$, which suggests that the integrated flux limited subsample is
complete. 

\vbox{
\vspace{-0.0cm}
\begin{center}
\leavevmode
\hbox{%
\epsfxsize=\hsize
\epsfbox[40 302 565 704]{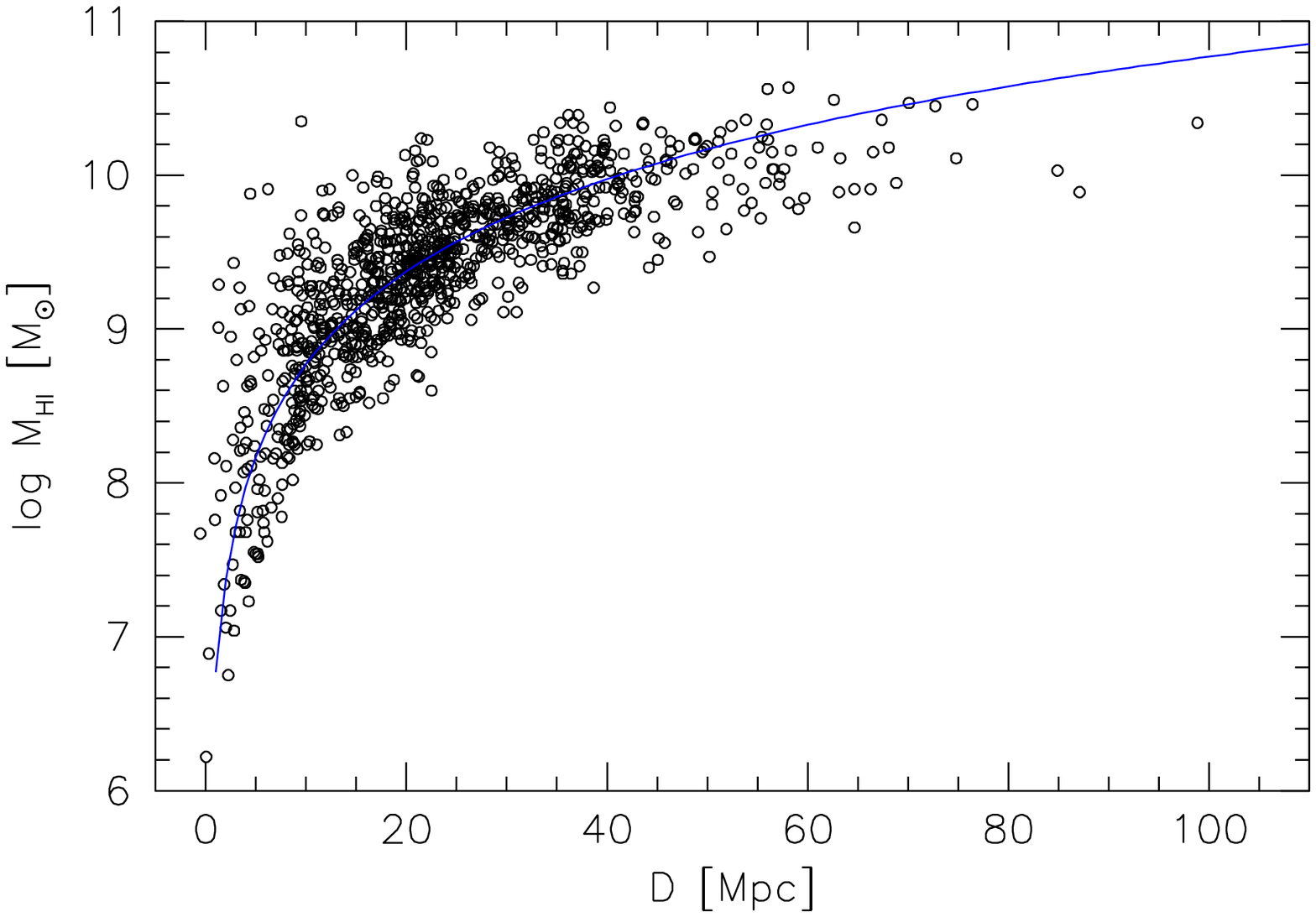}}
 \figcaption{\hi\ masses of BGC galaxies as a function
of  their distance. The solid line indicates an integrated flux limit of
$25 \,\rm Jy \, km \, s^{-1}$ above which the sample is `complete'.
 \label{mhi_D.fig}}
 \end{center}}

\section{Comparison to other surveys}
 In the previous section we have shown that the results on the \himf\
are internally consistent.  We now compare the results with those of
previous \hi\ surveys.  In Figure~\ref{compare.fig} we reproduce the best
fit Schechter function to the BGC \himf\ as a thick solid line.  In
addition we plot the \himfs\ from 
 the Arecibo \hi\ Strip Survey (AHISS, Zwaan et al. 1997, dotted line), 
 the Arecibo Slices (AS, Schneider et al. 1998), 
 the Arecibo Dual Beam Survey (ADBS, Rosenberg \& Schneider 2002, dashed
line)
 and the HIPASS South Celestial Cap (SCC, Kilborn 2000, long-dashed
line).  The ranges over which the Schechter function have been plotted
indicate the \hi\ mass bracket within which the \himfs\ have been reliable
determined from the different surveys.  All curves have been converted
to $H_0=75 \,\rm km\, s^{-1} Mpc^{-1}$. The Schechter parameters of the
various \himfs\ are given in Table~\ref{compare.tab}.

\begin{deluxetable*}{lcccl}
\tabletypesize{\scriptsize}
\tablecaption{Schechter parameters of \hi\ mass functions \label{compare.tab}}
\tablewidth{0pt}
\tablehead{
\colhead{Sample} &  
\colhead{$\alpha$} &  
\colhead{\mhis\ } &  
\colhead{$\theta^*$}  & 
\colhead{Reference} \\
\colhead{} &
\colhead{} &
\colhead{[$h_{75}^{-2} \msol$]} &
\colhead{[$10^{-4} h_{75}^3 \rm Mpc^{-3}$]} 
}  

\startdata
HIPASS BGC 	& -1.30	&  9.79 & 86 & this paper\\
AHISS		& -1.20	&  9.80 & 59 & Zwaan et al. (1997)\\
HIPASS SCC      & -1.52	&  10.1	& 32 & Kilborn (2000) \\
HIPASS HIZSS	& -1.51 &  9.70 & 60 & Henning et al. (2000)\\
ADBS		& -1.53 &  9.88 & 50 & Rosenberg \& Schneider (2002)\\
\enddata


\end{deluxetable*}

All curves agree very well at the high mass end, which can also be seen
from the value of $\log \mhis$, which is $\approx 9.8$ for all
surveys.  Discrepancies arise at the low mass end, where Zwaan et al. 
(1997) found $\alpha=-1.2$ and Rosenberg \& Schneider found
$\alpha=-1.53$.  Undoubtedly, part of these differences can be explained
by small number statistics, but different treatments of the survey
completeness and different \himf\ estimators may also cause discrepancies.

\vbox{
\vspace{-0.0cm}
\begin{center}
\leavevmode
\hbox{%
\epsfxsize=\hsize
\epsfbox[32 322 566 692]{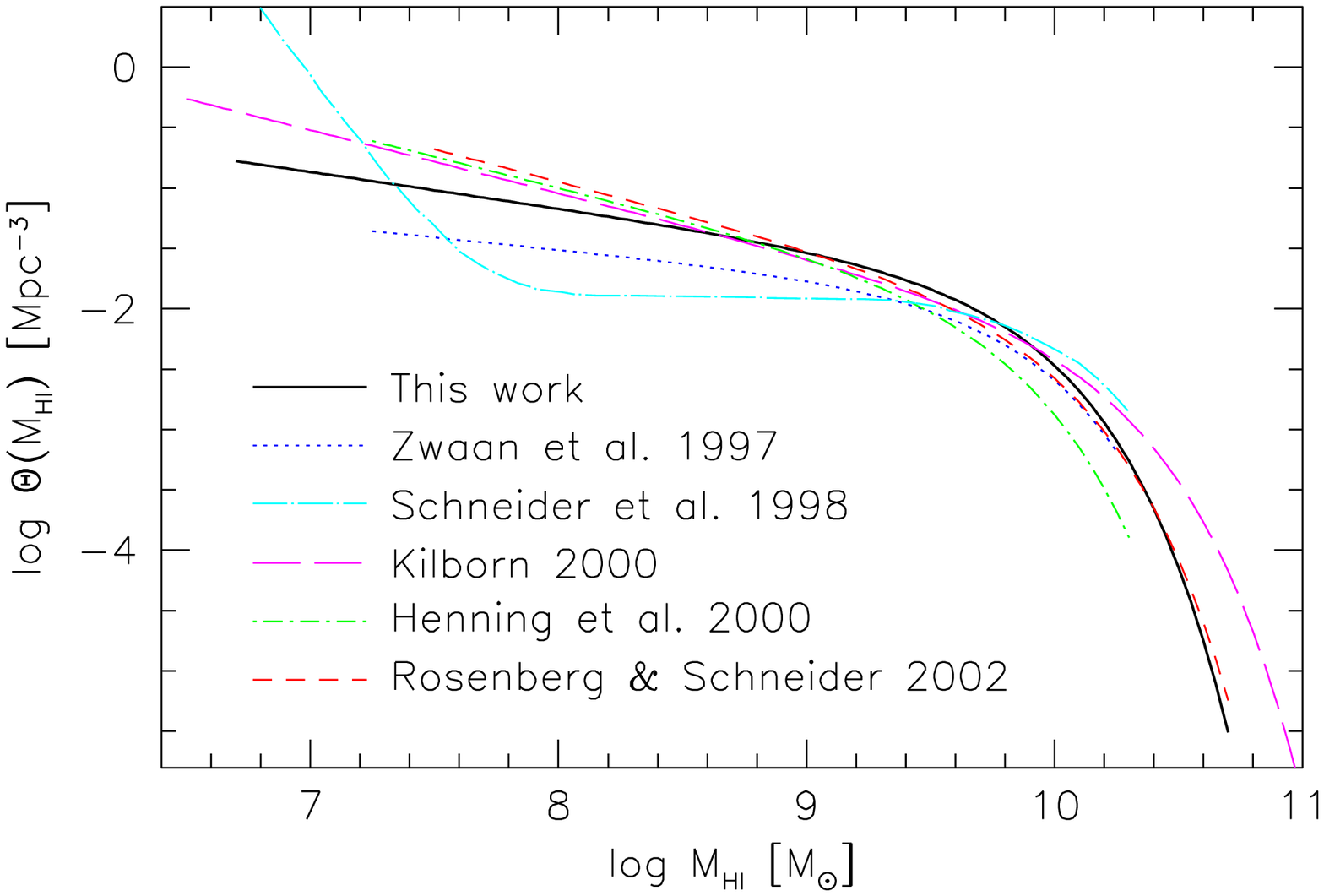}}
\figcaption{Comparison of the HIPASS BGC \himf\ with previous
calculations from blind \hi\ surveys.
Shown are \himfs\ from
 the AHISS (Zwaan et al. 1997), 
 the AS (Schneider et al. 1998),
 the HIPASS SCC (Kilborn 2000),
 the HIPASS HIZSS (Henning et al. 2000), 
 and the ADBS (Rosenberg \& Schneider 2002).
All mass functions have been converted to the same value of
$H_0=75~\hubble$ and are only plotted over the \hi\ mass range 
in which they have been measured. 
 \label{compare.fig}}
 \end{center}}

The AHISS (Zwaan et al.  1997) and the ADBS (Rosenberg \& Schneider
2002) are both drift-scan surveys performed with the Arecibo telescope. 
The AHISS is more sensitive with an $1\sigma$ rms noise of $\sim 0.75$
mJy/beam compared to $\sim 3.5$ mJy/beam per 32 \kms\ for the ADBS,
but the 
sky coverage of the ADBS was much larger which is reflected in the number of
detections: 66 in the AHISS and 265 in the ADBS.  Both galaxy samples
are based on by-eye examinations of the \hi\ spectra, which means that
the completeness limits have to be calculated {\em a posteriori}. 
Assessing the completeness limits of the drift-scan surveys is
complicated because the sensitivity is a function of declination offset,
which is particularly important if side-lobe detections have to be taken
into account (AHISS).  Zwaan et al.  (1997) derived an analytical
expression for the `detectability' of sources, Rosenberg \& Schneider
(2002) used a large number of synthetic sources to test their
completeness.  One of the differences in the analysis of both surveys is
the treatment of the variation of sensitivity $S$ with velocity width,
which can be expressed as $S\propto\Delta V^{\beta}$.  Assuming optimal
smoothing and uncorrelated noise, the expected value of $\beta$ is 0.5,
which was adopted by Zwaan et al.  (1997).  Rosenberg \& Schneider
(2002) argue that $\beta=0.75$ which leads to a reduced sensitivity to
large line width sources, and hence, via the Tully-Fisher relation, to a
reduced sensitivity to high \hi\ mass sources. 

With a peak flux limit of 116 mJy, the BGC is a relatively shallow
survey compared to the ADBS and the AHISS.  However, thanks to the BGC's
large sky coverage, the number counts at the low mass end are better
than those from AHISS and ADBS, and the BGC \himf\ extends to lower \hi\
masses than either of the Arecibo surveys.  This low sensitivity causes
all low mass galaxy detections to be at small distances (see
Figure~\ref{mhi_D.fig}), which implies that the Eddington effect might have
a significant influence on the \himf\ shape. 

Davies et al.  (2001) recently advocated an \himf\ slope of
$\alpha\approx -2$, which they derived by extracting \hi\ spectra from
the HIPASS public release data base, and compare the distance
distribution of detections with what would be expected for different
\himfs.  This method intrinsically assumes that a relationship exists
between distance and the minimal detectable \hi\ mass at that
distance.  We have argued in Section~\ref{methods.sec} that this
assumption does not hold for \hi\ selected galaxy samples.  The
validity of the Davies et al.  (2001) result is therefore unclear.

\section{Dependence of the \himf\ on galaxy type}
 The BGC was cross-correlated with the LEDA data base to find matches
with cataloged galaxies (see Jerjen et al.  2003).  In addition to
this, Ryan-Weber et al.  (2002) searched for uncataloged BGC galaxies on
the Digitized Sky Survey, and determined morphological types for 
these
new galaxies.  In total, morphological type information is now available
for 892 galaxies out of our total sample of 1000.  The 108 unclassified
galaxies consist mainly of galaxies at low Galactic latitude, groups or
pairs for which a unique match between \hi\ signal and optical galaxy
could not be made, or detections for which a cross correlation with LEDA
galaxies is otherwise ambiguous. 

In order to determine the morphological type dependence on the \himf\,
we divide the BGC into five subsets of galaxies: E-S0, Sa-Sb, Sbc-Sc,
Scd-Sd, and Sm-Irr.  Not surprisingly for an HI selected sample, the
subset of early-type galaxies contains only 43 galaxies, which is
marginally sufficient to calculate a meaningful \himf.  In addition, we
calculate the \himfs\ for galaxy samples divided into `late' and
'early', where galaxy types later than Sb are regarded as `late'. 

Figure~\ref{morph.fig} shows the \hi\ mass functions for different
morphological types.  We have chosen to apply the $1/V_{\rm max}$ method
here, because the subsamples are uncomfortably small for a 2-dimensional
analysis.  For reference, the best fit \himf\ for the total BGC
is indicated by a dotted line in each panel.  Best fit Schechter
functions to the individual \himfs\ are also shown, but note that for the
E-S0 subsample the normalization and \mhis\ are very poorly defined. 
All \himfs\ are multiplied by $1000/892$ to correct for the
incompleteness in morphological classification.  Table~\ref{morph.tab}
includes the best-fit Schechter parameters for all morphological types. 

There is no clear variation in the \himf\ slope between types Sa and Sd,
the \himf\ is flat with $\alpha\sim -1.0$.  Only for galaxies in the
Sm-Irr bin, we see a steepening of the \himf\ to $\alpha=-1.4$.  This
effect is very similar to what is found for optical luminosity functions
(Marzke et al.  1998).  Zwaan et al.  (2001) used the Marzke et al. 
(1998) luminosity functions and fitted relations between \hi\ mass and
optical luminosity to derive type specific \hi\ mass functions.  These
are in good agreement with what is found here, and also show a
steepening for the very late-type galaxies.  Apparently, both the number
density of low luminosity and low \hi\ mass galaxies are dominated by
late-type galaxies. 

Another conclusion from Figure~\ref{morph.fig} is that the characteristic
mass \mhis\ is much higher for types Sbc-Sc than for the other types. 
This reflects that most galaxies with high \hi\ masses are of types
Sbc-Sc, and that these types are the most dominant contributors to the
total \hi\ mass density. 

\vbox{
\vspace{-0.0cm}
\begin{center}
\leavevmode
\hbox{%
\epsfxsize=\hsize
\epsfbox[32 160 566 692]{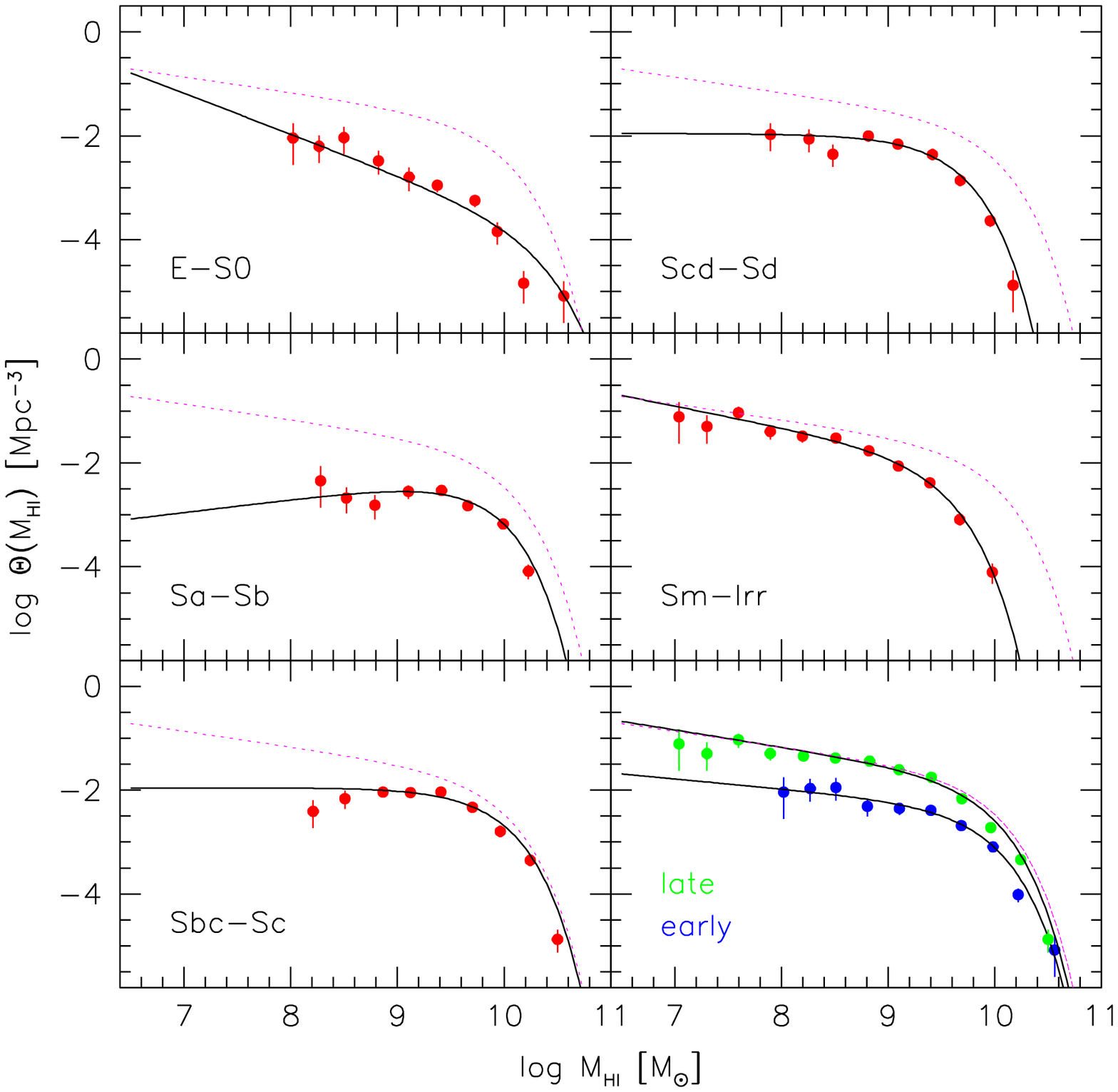}}
\figcaption{HI mass function for different morphological types. The solid
lines show Schechter fits to the type-specific \himfs\ and the dashed
line is \himf\ for the total sample. In the lower right panel, late-type
galaxies are all types later than Sb. 
\label{morph.fig}}
\end{center}}

\begin{deluxetable}{lcccl}
\tabletypesize{\scriptsize}
\tablecaption{Type-specific \hi\ mass functions \label{morph.tab}}
\tablewidth{0pt}
\tablehead{
\colhead{Type} &  
\colhead{$\alpha$} &  
\colhead{\mhis\ } &  
\colhead{$\theta^*$} \\
\colhead{} &
\colhead{} &
\colhead{[$h_{75}^{-2} \msol$]} &
\colhead{[$10^{-4} h_{75}^3 \rm Mpc^{-3}$]} 
}  
\startdata
E-S0 	& -1.78 & 10.2$^{\rm a}$  & 0.92$^{\rm a}$ \\
Sa-Sb	& -0.75 & 9.65 	& 22 \\
Sbc-Sc	& -1.00	& 9.77 	& 49 \\
Scd-Sd 	& -1.01 & 9.42 	& 46 \\
Sm-Irr	& -1.41 & 9.32 	& 61 \\
\tableline
late  	& -1.33 & 9.75 	& 78 \\
early	& -1.19 & 9.76	& 21 \\

\enddata
\tablenotetext{a}{Values uncertain because a Schechter function is a
poor fit to the data points}
\end{deluxetable}

Similarly, the \himf\ can be divided into galaxies of different optical
surface brightness.  The question of how much low surface brightness
(LSB) galaxies contribute to the local \hi\ density and the total local
baryon budget has been addressed before by several authors.  Based on
local galaxy samples with targeted 21cm spectroscopic follow-up, Briggs
(1997a, 1997b) concluded that LSB galaxies make a $\sim 10\%$ addition
to the total \hi\ mass density.  Zwaan, Briggs \& Sprayberry (2001) used
the AHISS to find that galaxies with central surface brightness $>23.0
\,\rm mag \, arcsec^2$ contribute $18\%$ to the \hi\ density.  The
luminosity density contained in LSB galaxies is also found to be low
(Sprayberry et al.  1997, Driver 1999, de Jong \& Lacey 2000, Zwaan et
al.  2001). 

Measurements of optical surface brightness for the BGC are drawn from
LEDA.  We choose to use the mean effective surface brightness
$\mu_{\rm eff}$, which is defined as the mean surface brightness
inside the aperture enclosing one-half the total light.
Unfortunately, the measurements are not complete: only for 600
galaxies in the BGC sample is $\mu_{\rm eff}$ available. The
uncertainties in the \himfs\ for LSB and high surface brightness (HSB)
galaxies are therefore large.  We divide the sample into two
subsamples, where we use $\mu_{\rm eff}=24.0 \,\rm mag\, arcsec^{-2}$
as the boundary between a LSB and a HSB galaxy. This separation
results in 96 LSB and 504 HSB galaxies.  For a exponential disk, the
value of $\mu_{\rm eff}=24.0 \,\rm mag\, arcsec^{-2}$ compares to a
central surface brightness of $\mu_{\rm eff}=22.2\,\rm mag\,
arcsec^{-2}$.  The values of $\mu_{\rm eff}$ are uncorrected for
inclination and dust extinction.

\vbox{
\vspace{-0.0cm}
\begin{center}
\leavevmode
\hbox{%
\epsfxsize=\hsize
\epsfbox[32 322 566 692]{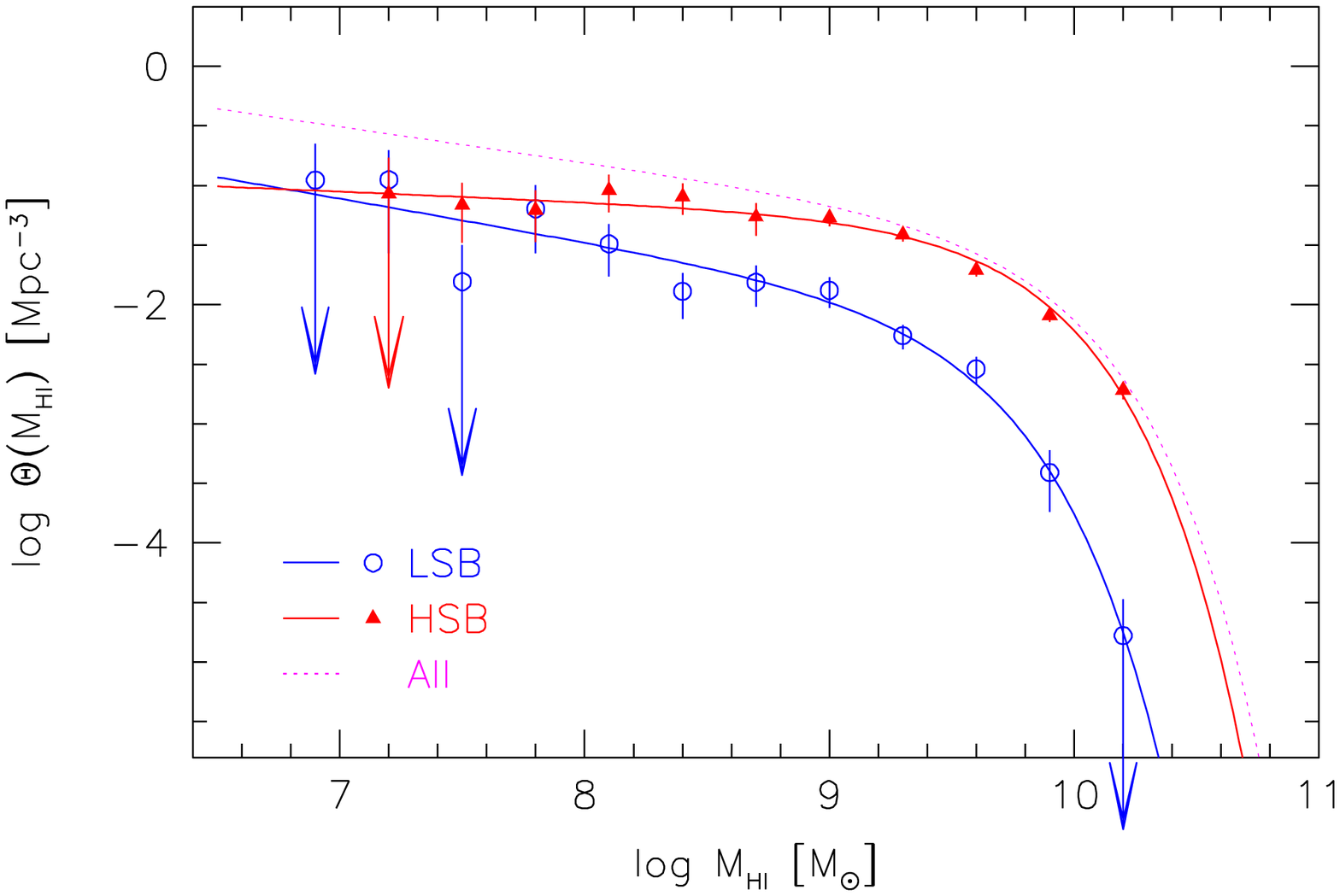}}
\figcaption{HI mass function for low surface brightness and high surface
brightness galaxies. The \himf\ for the total sample is drawn as a
dashed curve.
\label{surfbr.fig}}
\end{center}}

Figure~\ref{surfbr.fig} shows \himfs\ for the LSB and HSB subsamples. 
Both are corrected by a factor $1000/600$ to account for the
incompleteness in surface brightness measurements.  Similarly to what
was found by Briggs (1997a), we see that the \himf\ for LSB galaxies is
steeper ($\alpha=-1.35$) than that for HSB galaxies ($\alpha=-1.2$). 
The obvious reason for this is that there is a relation between surface
brightness and \hi\ mass, in the sense that LSB galaxies have lower
total \hi\ masses.  LSB galaxies therefore increasingly populate the
lower \hi\ mass bins, resulting in a steeper \hi\ mass function compared
to HSB galaxies.  Nonetheless, LSB galaxies contribute only very little
to the total \hi\ mass density.  Based on the Schechter fits we find
that this contribution is $\sim 15\%$. 

\section{The \hi\ mass density} \label{dens.sec}
 Figure~\ref{hidens.fig} shows the \hi\ mass density $\rho_{\rm HI}$
contained in galaxies of different \hi\ mass.  The heavy solid line
shows the converted best-fit Schechter function, and the dotted and
dashed lines show the \hi\ mass densities derived by Zwaan et al. 
(1997) and Rosenberg \& Schneider (2002), respectively.  The bivariate
\hi\ mass density distribution in the ($\mhi,W_{20}$) plane is shown in
Figure~\ref{2ddens.fig}.  This figure clearly shows that the gas mass
density is dominated by galaxies with \hi\ masses around \mhis\ and
velocity widths of $\sim 250$ \kms. 

\vbox{
\vspace{-0.0cm}
\begin{center}
\leavevmode
\hbox{%
\epsfxsize=\hsize
\epsfbox[32 310 566 692]{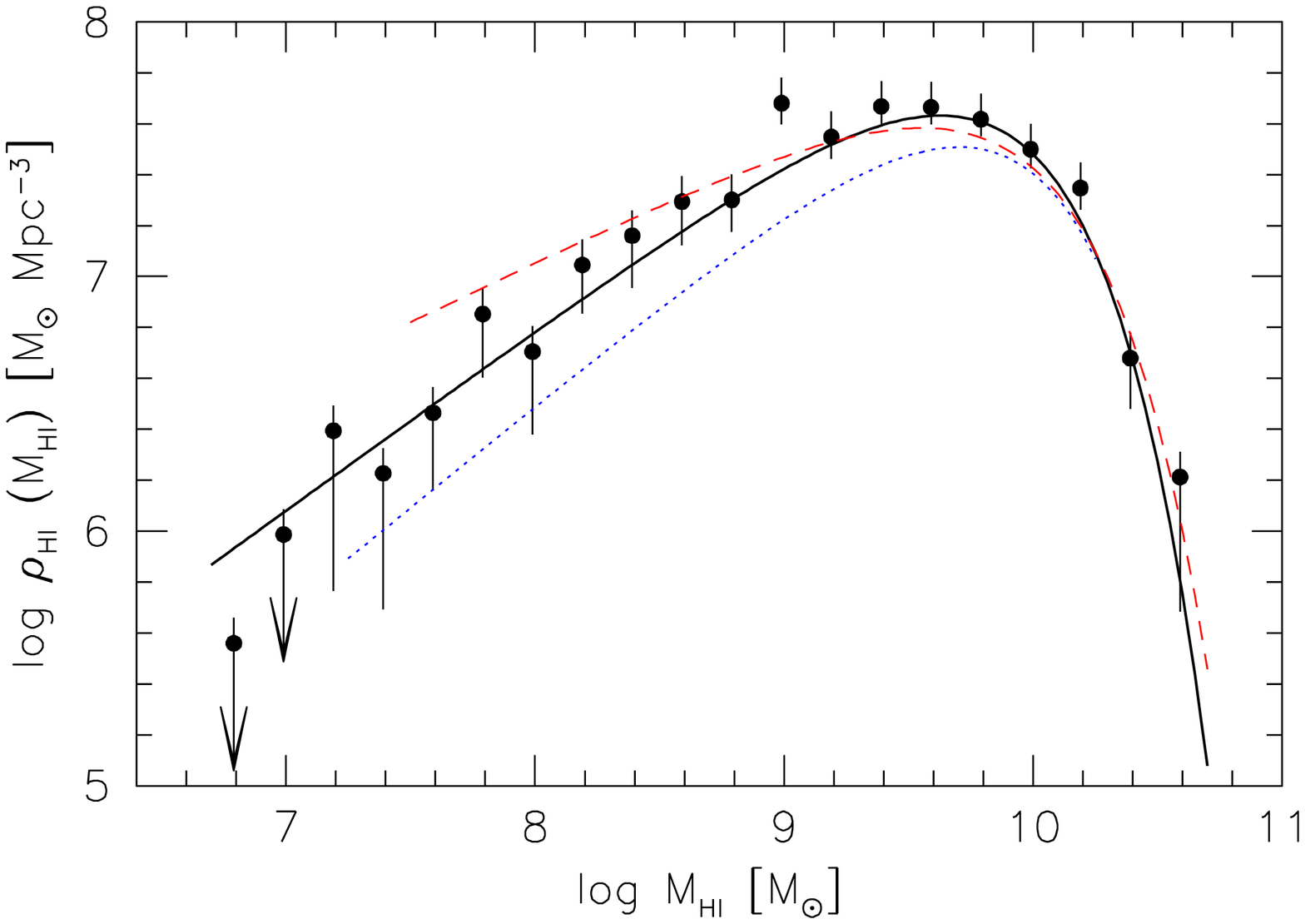}}
\figcaption{The distribution of \hi\ mass density over different \hi\
masses.  The points show the measured mass density derived by
multiplying the 2DSWML \himf\ by \mhi.  The solid line is the converted
best fit Schechter function.  The dotted line is the mass density
distribution function from Zwaan et al.  (1997) and the dashed line is
that derived by Rosenberg \& Schneider (2002). 
 \label{hidens.fig}}
 \end{center}}

The total \hi\ mass density contained by galaxies in the local
universe is calculated with $\rho_{\rm
HI}=\theta^*\Gamma(2+\alpha)\mhis$, and is found to be $\rho_{\rm
HI}=(6.9 \pm 1.1) \times 10^7 \, h_{75}\,\msol\,\rm Mpc^{-3}$.  A
straight summation of $\mhi\theta(\mhi)$ using the points in
Figure~\ref{himfswml2d.fig} gives a slightly lower value of $\rho_{\rm
HI}=(6.6 \pm 1.1)\times 10^7 \, h_{75}\,\msol\,\rm Mpc^{-3}$ because
this calculation does not include the extensions of the Schechter fit
beyond the region where we can measure it reliably.  Taking into
account all biases as summarized in Table~\ref{bias.tab}, we find
$\rho_{\rm HI}=(6.1 \pm 1.0) \times 10^7 \, h_{75}\, \msol\,\rm
Mpc^{-3}$ or $(4.1 \pm 0.7) \times 10^{-33} \rm g \,cm^{-3}$, where
the error is the $1\sigma$ uncertainty derived with the jackknife
method, which includes random errors and the effects of large scale
structure.  For comparison, Zwaan et al.  (1997) found $\rho_{\rm
HI}=4.3 \times 10^7 \, h_{75}\,\msol\,\rm Mpc^{-3}$ and using
Rosenberg \& Schneider's (2002) Schechter parameters we find
$\rho_{\rm HI}=7.1 \times 10^7 \, h_{75}\,\msol\,\rm Mpc^{-3}$. Briggs
(1990) used luminosity functions and a conversion factor from
luminosity to \hi\ mass to find $\rho_{\rm HI}=7.0 \times 10^7 \,
h_{75}\,\msol\,\rm Mpc^{-3}$, a value very close to our measurement.

Converting $\rho_{\rm HI}$ to the more convenient $\Omega_{\rm HI}$, the \hi\
mass density as a fraction of the critical density of the universe, we
find $\Omega_{\rm HI}=(3.8 \pm 0.6) \times 10^{-4} h_{75}^{-1}$ and after
making a correction for the 24\% helium mass fraction we find
$\Omega_{\rm atomic}=(4.8 \pm 0.8) \times 10^{-4} h_{75}^{-1}$.  This
measurement of $\Omega_{\rm HI}$ is very robust because the galaxies
that contribute most to the gas density are also the ones that dominate
the counting statistics ($\mhi=10^9 - 10^{10}~\msol$).  Uncertainties
in the faint-end slope $\alpha$ therefore contribute little to the total
error in $\Omega_{\rm HI}$ .  Furthermore, since these galaxies are
found at large distances, and therefore over a large region of the sky,
the effects of large scale structure are unimportant.

\vbox{
\vspace{-0.0cm}
\begin{center}
\leavevmode
\hbox{%
\epsfxsize=\hsize
\epsfbox[19 143 569 695]{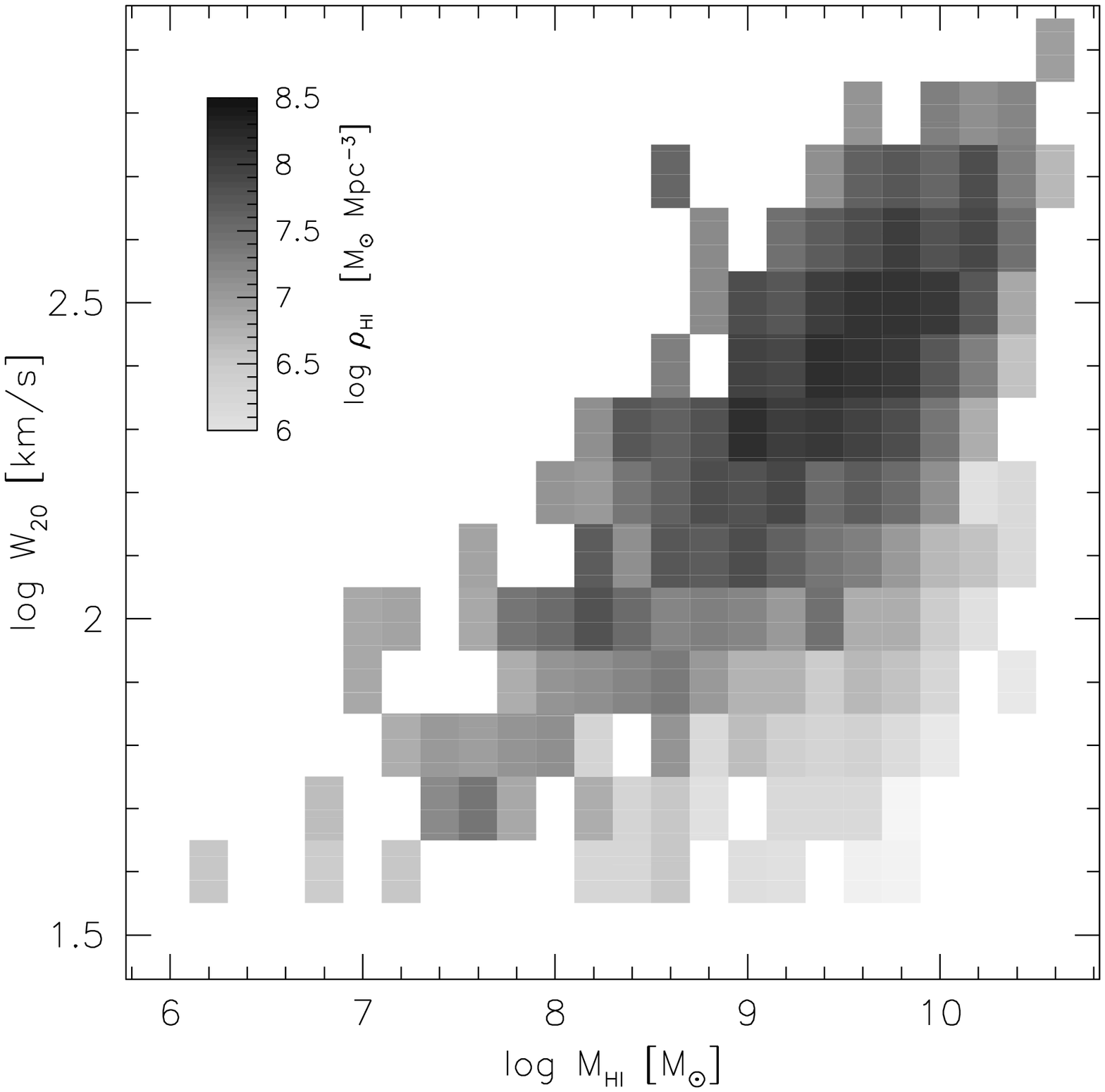}}
 \figcaption{Bivariate \hi\ mass density distribution in the ($\mhi,w_{20}$)
plane.  The greyscales are on a logarithmic scale and represent the mass
density in $\msol \rm Mpc^{-3}$ per decade of \mhi\ and decade of $w_{20}$.
The \hi\ mass density is dominated
by high mass, high velocity width galaxies. 
\label{2ddens.fig}}
\end{center}}

Fukugita, Hogan \& Peebles (1998) estimated that the mass fraction in
molecular gas is approximately 80\% of that of atomic gas.  Keres, Yun
\& Young (2002) measure the CO luminosity function and derive that the
mass fraction in molecular gas is approximately 60\%.  Averaging these
two values, we find that the total mass density in cool gas in the local
universe is $\Omega_{\rm cool\, gas}=7.6 \times 10^{-4} h_{75}^{-1}$. 
To put these numbers into perspective, the most recent determination of
the baryon density via the primeval deuterium abundance is $\Omega_{\rm
baryon}h_{75}^2 = 0.035 \pm 0.004$ (Burles, Nollett \& Turner 2001). 
Microwave background anisotropy measurements give slightly higher values
(e.g., de Bernardis et al.  2000).  We estimate from this that cool gas
in galaxies makes up approximately 2\% of the total baryon density in
the local universe.  The total mass in diffuse ionized intergalactic gas
is much higher than this.  From low redshift HST spectra Penton, Shull,
\& Stocke (2000) derived that the mass density in the Ly$\alpha$ forest
is approximately 20\% of $\Omega_{\rm baryon}$. 

Our derived value of $\Omega_{\rm HI}$ is approximately 5 times lower
than that at redshifts $2-4$ (Ellison et al.  2001, P{\'e}roux et al. 
2001, Storrie-Lombardi, McMahon, \& Irwin 1996).  A gradual conversion
from neutral gas to stars in the disks of galaxies is generally believed
to cause this decline in $\Omega_{\rm HI}$ (Lanzetta et al.
1995, Pei, Fall \& Hauser 1999), but recent results on
high column density QSO absorption-line systems at intermediate redshifts
have confused this picture.  Lane (2000) searched for intermediate $z$
21cm absorption in Mg{\sc ii} selected systems, and used Mg{\sc ii}
statistics to bootstrap the $\Omega_{\rm HI}$ values.  The same
statistics was used by Rao \& Turnshek (2000) for a Ly-$\alpha$ survey
of $z<1.65$ Mg{\sc ii} selected systems.  Both authors find values of
$\Omega_{\rm HI}(z<1.65)$ consistent with those at higher redshifts,
indicating that the neutral gas density does not evolve strongly from
high $z$ to the present.  Churchill (2001) recently performed an
unbiased survey for low-$z$ Mg{\sc ii} systems and used the same
statistics for Mg{\sc ii} systems as Rao \& Turnshek did, and derived
$\Omega_{\rm HI}(z=0.05)$, which is a factor 5 larger than our results. 
The uncertainties on all these measurements are very large as they
suffer from small number statistics.  The evolution of $\Omega_{\rm HI}$
from $z=2$ to the present time therefore remains very uncertain and more
intermediate redshift optical and 21cm surveys are required to constrain
the evolution of \hi. 

\vbox{
\vspace{-0.0cm}
\begin{center}
\leavevmode
\hbox{%
\epsfxsize=\hsize
\epsfbox[32 160 576 692]{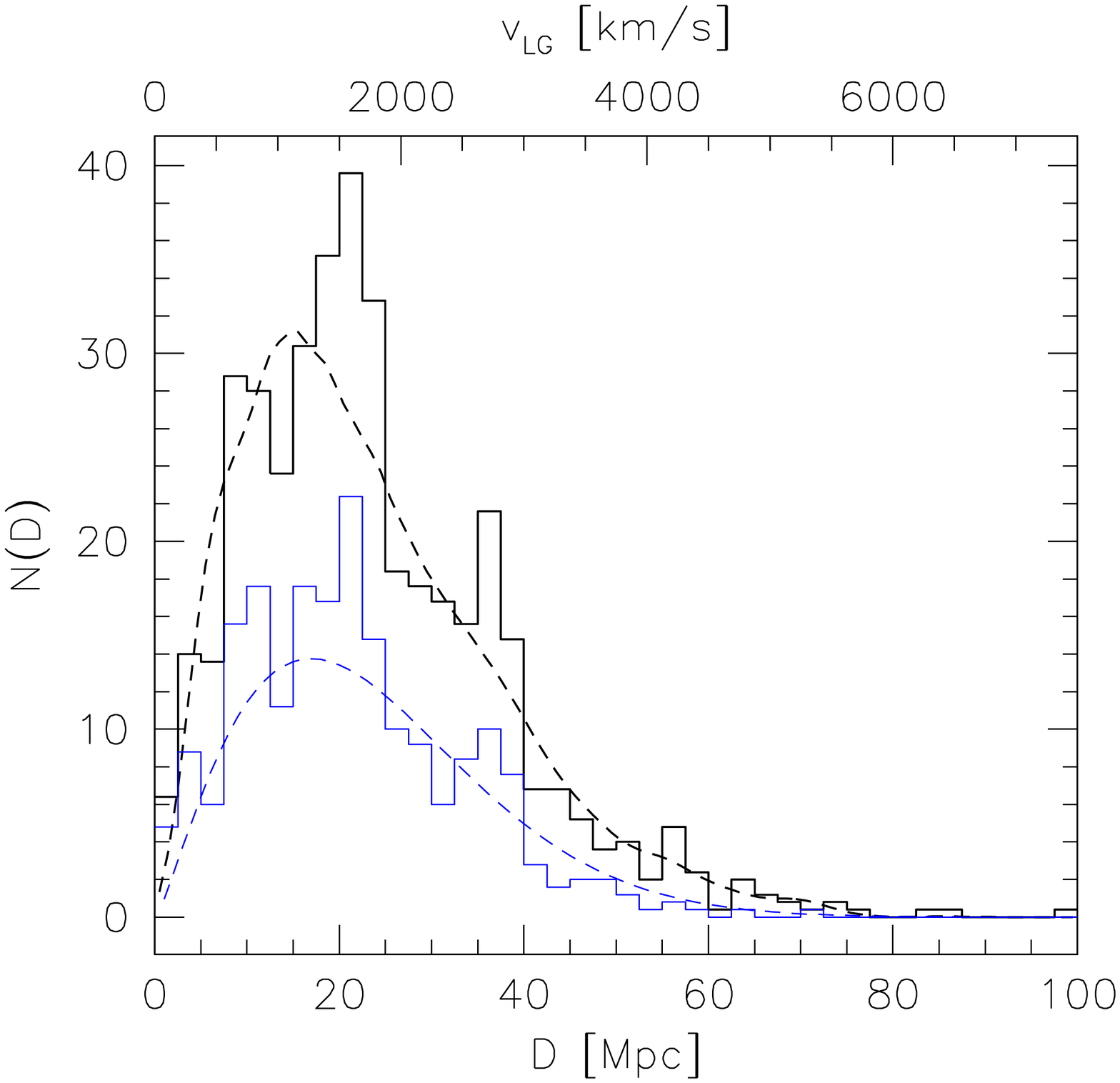}}
 \figcaption{Distance distribution of the BGC.  The thick histogram shows
the measured redshift distribution of galaxies in the BGC, the smooth
dashed curve is the predicted distribution based on the selection
function $S(D)$, which is calculated with the 2DSWML method.
Both distributions are normalized to $\Delta D=1\, \rm Mpc$.  The thin
lines show the same for a subsample with $S_{\rm int}>25\, \rm Jy\, km\, s^{-1}$. 
\label{selec.fig}}
\end{center}}

\section{The BGC Selection Function}
 As described in Section~\ref{selec.sec}, the 2DSWML method allows us
the calculate the selection function of the BGC.  The selection function
is an important input into calculations of clustering in the BGC, like
the 2-point correlation function (Meyer et al.  2003b).  It is
therefore interesting to test whether the calculated selection function
$S(D)$ is a valid approximation of the true selection of the survey.

In Figure~\ref{selec.fig} we compare the measured redshift distribution of BGC
galaxies with the distribution implied by the derived selection
function.  
 The overall redshift distribution agrees well with the calculated
smooth curve.  This is also reflected in the predicted number counts of
galaxies from the selection function, which is 0.0456
galaxies/degree$^2$.  This is only 6\% lower than the measured number
count in the BGC of 0.0485 galaxies/degree$^2$. This implies that if we
had used the number counts to normalize the \himf\, similar to what is
occasionally done for optical LFs (e.g., Norberg et al. 2001), we would
have found essentially the same normalization. 

By comparing $N(D)$ with the smooth curve, two overdensities become
apparent: one at $\sim 22$ Mpc and one at $\sim 36$ Mpc. See Koribalski
et al. (2003) for a detailed discussion of the large-scale structure in
the BGC.
Also shown in Figure~\ref{selec.fig} is the redshift distribution of the
integrated flux limited subsample, and the predicted distribution
calculated with Eq.~\ref{selec.eq}.  The same density structure can be
seen here, although less pronounced than for the full galaxy sample.

\section{Conclusions}
 We have used the HIPASS Bright Galaxy Catalog (BGC, Koribalski et
al. 2003) consisting of the 1000 southern galaxies with highest peak
flux densities, to measure the \hi\ mass function (\himf) of galaxies
in the local universe.  This is the largest sample of galaxies ever
used to measure the \himf, and contains galaxies over the \hi\ mass
range from $\log (\mhi/\msol)+2\log h_{75}=6.8$ to $10.6$.  
We develop a bivariate
stepwise maximum likelihood method to measure the \himf. This method
solves for \hi\ mass and velocity width, and then integrates over
velocity width to find the \himf. We show that this method is a
reliable estimator and insensitive to the effects of large scale
structure.  The resulting \himf\ can be well fit with a Schechter
function with parameters $\alpha=-1.30 \pm 0.08$, $\log
(\mhis/\msol)=9.79 \pm 0.06 \,h^{-2}_{75}$, and $\theta^*=(8.6 \pm
2.1)\times 10^{-3} \,h^{3}_{75}\,{\rm Mpc}^{-3}$. 
We find that the
faint-end slope of the \himf\ is dependent on morphological
type. Late-type galaxies show the steepest faint-end slopes and these
galaxies dominate the statistics at low \hi\ masses.  We extensively
test the influence of possible biases in the \himf\ determination,
including peculiar motions of galaxies, inclination effects, selection
biases and large scale structure, and quantify these biases.  The
integral \hi\ mass density in the local universe is found to be
$\rho_{\rm HI}=(6.1 \pm 1.0) \times 10^7 \, h_{75}\, \msol\,\rm
Mpc^{-3}$, contributing a fraction $\Omega_{\rm HI}=(3.8 \pm 0.6)
\times 10^{-4} h_{75}^{-1}$ of the critical density of the universe.

\acknowledgments


\begin{references}

\reference{} Babul, A. \& Ferguson, H.~C. 1996, \apj, 458, 100

\reference{} Babul, A. \& Rees, M.~J. 1992, \mnras, 255, 346

\reference{} Barkana, R.~\&
Loeb, A.\ 1999, \apj, 523, 54 

\reference{} Barnes, D.~G.~et al.\
2001, \mnras, 322, 486 

\reference{} Blanton,
M.~R.~et al.\ 2001, \aj, 121, 2358 

\reference{} Bottinelli, L., Gouguenheim, L., 
Fouque, P., \& Paturel, G.\ 1990, \aaps, 82, 391 

\reference{} Braun, R.\ 1997, \apj, 484,
  637

\reference{}Briggs, F.~H. 1990, \aj, 100, 999

\reference{}Briggs, F.~H., \& Rao, S.  1993, \apj, 417, 494

\reference{} Briggs, F.~H.\ 1997a, \apj,
484, 618 

\reference{} Briggs, F.~H.\ 1997b, \apjl, 
484, L29 

\reference{} Burles,
S., Nollett, K.~M., \& Turner, M.~S.\ 2001, \apjl, 552, L1 

\reference{} Churchill, C.~W.\ 2001, 
\apj, 560, 92 

\reference{} Corbelli, E., Salpeter, E.~E., Bandiera, R., 2001, \apj,
  550, 26

\reference{} Davis, M.~\&
Huchra,  J.\ 1982, \apj, 254, 437 

\reference{} Davies, J.~I., de
Blok, W.~J.~G., Smith, R.~M., Kambas, A
Sabatini, S., Linder, S.~M., \& Salehi-Reyhani, S.~A. 2001, \mnras,
328, 1151 

\reference{} de Bernardis, 
P.~et al.\ 2000, \nat, 404, 955 

\reference{} de Jong, R.~S.~\& 
Lacey, C.\ 2000, \apj, 545, 781 

\reference{} Dekel, A.~\& Silk,
J.\ 1986, \apj, 303, 39 

\reference{} Dickey, J.~M., Mebold, U.,
Stanimirovic, S., \& Staveley-Smith, L.\ 2000, \apj, 536, 756

\reference{} Driver, S.~P. 1999, \apj, 526, L69

\reference{} Efstathiou, G., Ellis, R.~S., \& Peterson, B.~A.  1988,
        \mnras, 231, 479

\reference{} Ellison, S.~L., Yan,
L., Hook, I.~M., Pettini, M., Wall, J.~V., \& Shaver, P.\ 2001, \aap,
379,  393 

\reference{} Folkes, S.~et al.\
1999,  \mnras, 308, 459 

\reference{}
Fukugita,  M., Hogan, C.~J., \& Peebles, P.~J.~E.\ 1998, \apj, 503, 518 

\reference{} 
Gibson, S.~J., Taylor, A.~R., Higgs, L.~A., \& Dewdney, P.~E.\ 2000,
\apj, 540, 851 

\reference{} Haynes, M.~P. \& Giovanelli, R.  1984, \aj, 89, 758

\reference{} Huizinga, 
J.~E.~\& van Albada, T.~S.\ 1992, \mnras, 254, 677 

\reference{} Henning, P.~A.~et
al.\  2000, \aj, 119, 2686 

\reference{} Hoyle, F.~\&
Vogeley, M.~S.\ 2002, \apj, 566, 641 

\reference{} Jerjen, H. et al. 2003, in preparation

\reference{} Jing, Y.~P., B{\" o}rner, G., \& Suto, Y.\ 2002, \apj, 564, 15 

\reference{}Keres, D., Yun, M.,~S., \& Young, J.~S.\ 2002, \apj, in press

\reference{}Kilborn, V.~A. 2000, Ph.D. Thesis, University of Melbourne

\reference{} Kilborn, V.~A.~et
al.\ 2002, \aj, 124, 690 

\reference{} Knapp, G.~R.\ 1974, \aj, 79, 527 

\reference{}Koribalski, B.~S. et al. 2003, submitted to AJ

\reference{} Kraan-Korteweg, 
R.~C., van Driel, W., Briggs, F., Binggeli, B., \& Mostefaoui, T.~I.\
1999, \aaps, 135, 255 

\reference{}Lane, W.~M. 2000, Ph.D. Thesis, University of Groningen

\reference{}Lang, R.~H. et al.\ 2002, \mnras, in press

\reference{}
Lanzetta,  K.~M., Wolfe, A.~M., \& Turnshek, D.~A.\ 1995, \apj, 440, 435 

\reference{} Loveday, J.\ 2000, \mnras, 312, 557

\reference{} Lupton, R. 1993, Statistics in Theory and Practise
(Princeton: Princeton Univ. Press)

\reference{} Marzke, R.~O., da Costa, L.~N., Pellegrini, P.~S., Willmer,
        C.~N.~A. \& Geller, M.~J.\ 1998, \apj, 503, 617

\reference{} Meyer, M.~J., Zwaan, M.~A., Webster, R.~L., et al. 2003a, in prep.

\reference{} Meyer, M.~J. et al.  2003b, in prep.

\reference{} Norberg, P. et al. 2001, astro/ph-0111011

\reference{} Pei, Y.~C.,
Fall,  S.~M., \& Hauser, M.~G.\ 1999, \apj, 522, 604 

\reference{} Penton, 
S.~V., Shull, J.~M., \& Stocke, J.~T.\ 2000, \apj, 544, 150 

\reference{} P{\' e}roux, C.~., Irwin,
M.~J., McMahon, R.~G., \& Storrie-Lombardi, L.~J.\ 2001, Astrophysics
and Space Science Supplement, 277, 551

\reference{}Plionis, M. \& Basilakos, S.\ 2002, \mnras, 330, 399

\reference{}Rao, S.~M. \& Turnshek, D.~A. 2000, ApJS, 130, 1

\reference{} Roberts, M.~S.~\&
Haynes, M.~P.\ 1994, \araa, 32, 115 

\reference{} Rosenberg, 
J.~L.~\& Schneider, S.~E.\ 2002, \apj, 567, 247 

\reference{} Rosenberg, 
J.~L.~\& Schneider, S.~E.\ 2000, \apjs, 130, 177 

\reference{} Ryan-Weber, E.~et
al.\ 2002, \aj, 124, 1954 

\reference{} Ryder, S.~D. et al.\ 2001, \apj, 555, 232

\reference{} Sandage, A., Tammann, G.~A., \& Yahil, A. 1979, \apj, 232, 352 

\reference{} Schechter, P.  1976, \apj, 203, 297

\reference{} 
Schlegel, D., Davis, M., Summers, F., \& Holtzman, J.~A.\ 1994, \apj,
427, 527 

\reference{} Schmidt, M.  1968, \apj, 151, 393

\reference{} 
Schneider, S.~E., Spitzak, J.~G., \& Rosenberg, J.~L.\ 1998, \apjl, 507,
L9 

\reference{} Sheth, R.~K.~\&
Diaferio, A.\ 2001, \mnras, 322, 901 

\reference{}
Staveley-Smith, 
L.~et al.\ 1996, Publications of the Astronomical Society of Australia,
13, 243 

\reference{}
Sprayberry, D., Impey, C.~D., Irwin, M.~J., \& Bothun, G.~D.\ 1997,
\apj, 482, 104 

\reference{}
Storrie-Lombardi, L.~J., McMahon, R.~G., \& Irwin, M.~J.\ 1996, \mnras, 
283, L79 

\reference{} 
Storrie-Lombardi, L.~J.~\& Wolfe, A.~M.\ 2000, \apj, 543, 552 

\reference{} Strauss, 
M.~A., Ostriker, J.~P., \& Cen, R.\ 1998, \apj, 494, 20 

\reference{} Tully, R.~B., \& Fisher, J.~R., 1977, \aap, 54, 661

\reference{} Verheijen, 
M.~A.~W.~\& Sancisi, R.\ 2001, \aap, 370, 765 

\reference{} 
Willick, J.~A., Strauss, M.~A., Dekel, A., \& Kolatt, T.\ 1997, \apj,
486,  629 

\reference{} Willmer, C.~N.~A. 1997, \aj, 114, 898

\reference{} Zwaan,
M.~A., Briggs, F.~H., \& Sprayberry, D.\ 2001, \mnras, 327, 1249 

\reference{}
Zwaan, M.~A., Briggs, F.~H., Sprayberry, D., \& Sorar, E.\ 1997, \apj,
490, 173 

\reference{} Zwaan,
M.~A., Verheijen, M.~A.~W., \& Briggs, F.~H.\ 1999, Publications of the
Astronomical Society of Australia, 16, 100

\end{references}
\end{document}